\documentclass[manuscript,review=false,anonymous=false]{acmart}

\usepackage{xcolor}
\usepackage{xspace}
\usepackage{colortbl}

\usepackage{booktabs}
\usepackage[table]{xcolor}
\usepackage{subcaption}
\usepackage{tabularx}
\usepackage{array}

\usepackage{listings}
\newcommand{\issuesql}{buggy SQL query\xspace}
\newcommand{\nlquery}{task requirements\xspace}

\lstdefinelanguage{SQL}{
    keywords={SELECT,FROM,WHERE,JOIN,LEFT,ON,AS,AND,OR,NOT,EXISTS,
              DISTINCT,WITH,RECURSIVE,UNION,ALL,CASE,WHEN,THEN,ELSE,
              END,GROUP,BY,MIN,GREATEST,IN},
    sensitive=false,
    morecomment=[l]{--},
    morestring=[b]',
}

\begin{document}

\title{How Do People Debug Complex SQL Queries With LLMs? A Qualitative Study}


\author{Nicole Sullivan}
\email{nsulliv@cs.washington.edu}
\affiliation{
  \institution{University of Washington}
  \city{Seattle}
  \state{Washington}
  \country{USA}
}

\author{Ella Cao}
\email{ellacao@cs.washington.edu}
\affiliation{
  \institution{University of Washington}
  \city{Seattle}
  \state{Washington}
  \country{USA}
}

\author{Leilani Battle}
\email{leibatt@cs.washington.edu }
\affiliation{
 \institution{University of Washington}
 \city{Seattle}
 \state{Washington}
 \country{USA}
}

\author{Magdalena Balazinska}
\email{magda@cs.washington.edu}
\affiliation{
 \institution{University of Washington}
 \city{Seattle}
 \state{Washington}
 \country{USA}
}



\begin{abstract}
While LLMs are increasingly used in various contexts, we lack a fundamental understanding of how they impact SQL query debugging. In response, we conducted a user study with 16 participants of varying SQL expertise, examining how they debug complex SQL queries with and without LLM support. Initial benchmarking indicated that our selected models are already capable SQL debuggers. Qualitative analyses of our study data revealed that while participants trusted the LLM, they varied widely in how and whether they delegated debugging tasks to the LLMs, which we characterized through a new three-state model of participants' SQL reasoning processes. Additionally, participants enforced specific collaborative roles for the LLM such as TA or sounding board, which participants controlled by withholding context from the LLM. These findings suggest that LLM-based tools could enable preset role selection, broader context management, and flexible automation features, including low automation options, to support SQL debugging moving forward.

\end{abstract}

\ccsdesc[500]{Human-centered computing~Empirical studies in HCI}
\ccsdesc[500]{Software and its engineering~Software testing and debugging}
\ccsdesc[500]{Information systems~Structured Query Language}



\settopmatter{printacmref=false} 
\renewcommand\footnotetextcopyrightpermission[1]{} 


\maketitle
\thispagestyle{empty} 
\fancyfoot{} 

\begin{sloppypar}

\newcommand{\StudyOverview}{
    \begin{figure*}[t]
      \centering
      \includegraphics[width=0.85\textwidth, alt={Diagram of a 90 minute study, pre survey, warm up, four debugging tasks with varying tool access, and post survey}]{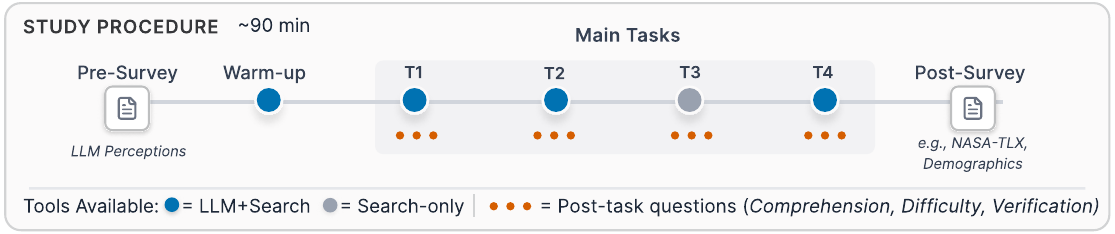}
      \caption{Participants completed a pre-survey and warm-up task, followed by four debugging tasks. The study concluded with a post-survey and optional interview (not shown).}
      \label{fig:study}
      \Description{A sample of tool availability varied across the debugging tasks. In this example, the warm up and Tasks 1, 2, and 4 provided a large language model plus search, while Task 3 provided search only.}
    \end{figure*}
}

\newcommand{\TaskExamples}{
    \begin{figure}[t]
      \centering
      \includegraphics[width=\textwidth, alt={Two SQL debugging examples linking task requirements to buggy SQL for quantification and recursion concepts.}]{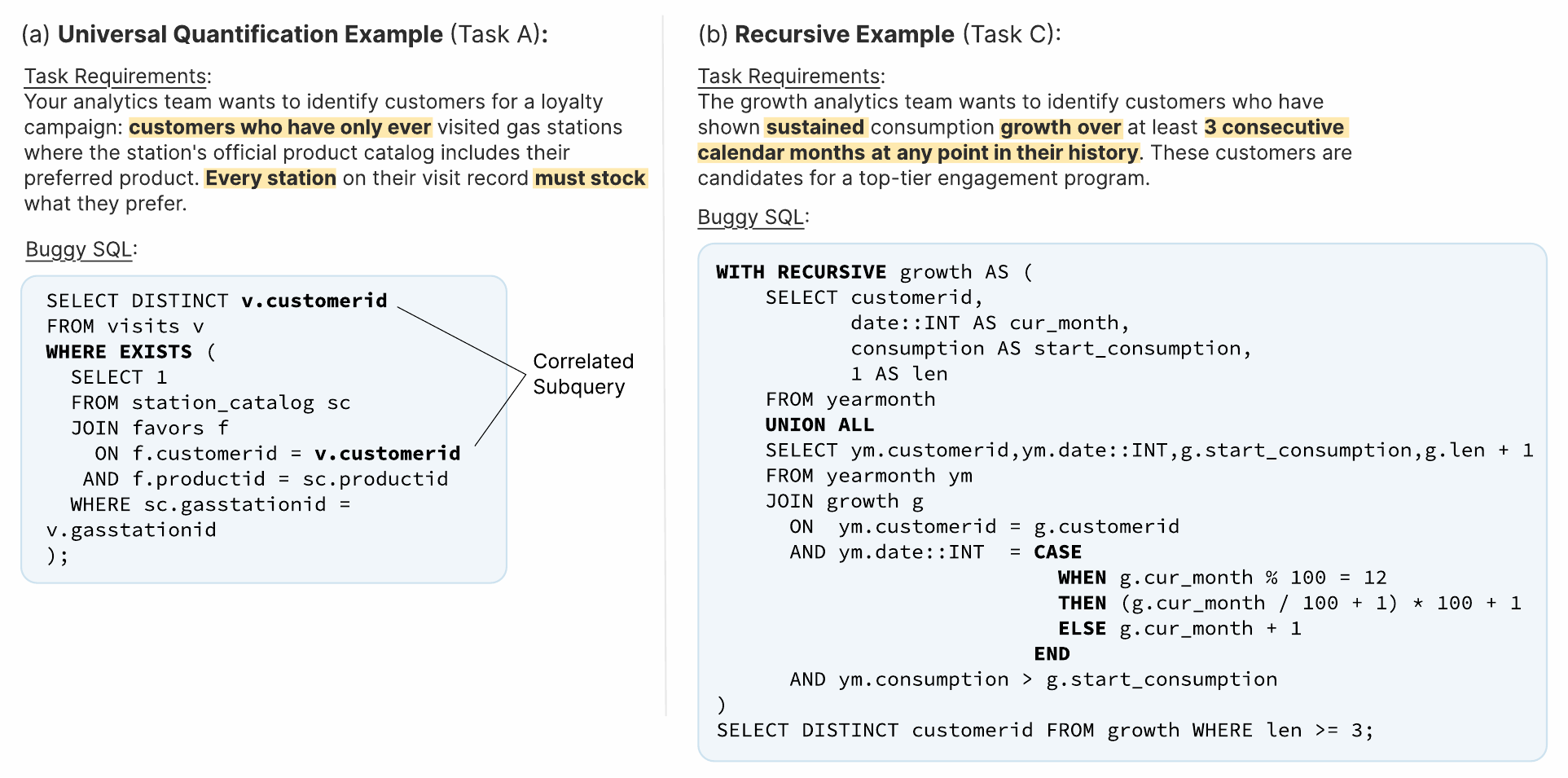}
      \caption{Representative universal-quantification (a) and recursive (b) debugging tasks. Highlighted text marks key task requirements; bold text marks complex SQL constructs in the buggy query.}
      \label{fig:task_examples}
      \Description{Two SQL debugging tasks, each with natural-language requirements and a buggy SQL query. Panel A shows a universal quantification task, with only, every, and must highlighted as key requirement components. Correlated subquery constructs are emphasized in the buggy SQL. Panel B shows a recursive task, with sustained growth over at least three consecutive months highlighted as a key requirement component. Recursive constructs and the case when statement are emphasized in the buggy SQL.}
    \end{figure}

}

\newcommand{\ParticipantInfo}{

\begin{table*}[t]
\centering
\caption{Participant demographics, SQL experience, self-reported LLM usage, and GPT model used in the study.}
\label{tab:participant-demographics}
\Description{A table with sixteen rows, one per participant labeled P01 through P16,
with columns for Age, Gender, Education Group, Experience, LLM Usage,
and GPT Model. Ages range from 18 to 24 up through 55 to 64, with
eleven of the sixteen participants between ages 18 and 34. Eleven
participants are male and five are female. Education group includes
five undergraduates, five graduate students, and six professionals.
Experience is rated Beginner for eight participants, Intermediate for
four participants, and Advanced for four participants. LLM usage is
Daily for eleven participants, with the rest using large language
models weekly, a few times a week, or rarely. Participants P01
through P13 used GPT model five point four, while participants P14 through P16 used
GPT model five point five.}
\small

\begin{tabularx}{\textwidth}{
    l |
    c
    c |
    l
    l |
    >{\raggedright\arraybackslash}X |
    l
}
\toprule
\textbf{Participant} &
\textbf{Age} &
\textbf{Gender} &
\textbf{Education Group} &
\textbf{Experience} &
\textbf{LLM Usage} &
\textbf{GPT Model} \\
\midrule

P01 & 18--24 & Male   & Undergrad    & Intermediate & Daily                      & GPT-5.4 \\
P02 & 18--24 & Male   & Undergrad    & Beginner     & 3--4 times/week             & GPT-5.4 \\
P03 & 18--24 & Female & Undergrad    & Beginner     & Weekly                      & GPT-5.4 \\
P04 & 18--24 & Female & Undergrad    & Beginner     & Daily                       & GPT-5.4 \\
P05 & 18--24 & Male   & Professional & Intermediate & Daily                       & GPT-5.4 \\
P06 & 18--24 & Male   & Grad student & Beginner     & Daily                       & GPT-5.4 \\
P07 & 18--24 & Male   & Grad student & Beginner     & Daily                       & GPT-5.4 \\
P08 & 25--34 & Male   & Grad student & Beginner     & Daily                       & GPT-5.4 \\
P09 & 35--44 & Female & Professional & Advanced     & Daily                       & GPT-5.4 \\
P10 & 25--34 & Female & Grad student & Beginner     & Weekly                      & GPT-5.4 \\
P11 & 55--64 & Male   & Professional & Advanced     & Daily                       & GPT-5.4 \\
P12 & 35--44 & Female & Professional & Intermediate & Daily                       & GPT-5.4 \\
P13 & 25--34 & Male   & Professional & Intermediate & Daily                       & GPT-5.4 \\
P14 & 18--24 & Male   & Undergrad    & Beginner     & Weekly                      & GPT-5.5 \\
P15 & 45--54 & Male   & Professional & Advanced     & Daily                       & GPT-5.5 \\
P16 & 35--44 & Male   & Grad student & Advanced     & Rarely (only when required) & GPT-5.5 \\
\bottomrule
\end{tabularx}
\end{table*}
}


\newcommand{\OutcomesTable}{
\begin{table}[t]
    \centering
    \caption{Task completion, solution correctness, and comprehension accuracy by access condition. Percentages are calculated using the denominator shown in each cell.}
    \label{tab:task-outcomes}
    \Description{A table with three columns, Search-only, LLM+Search, and Overall, and three sections: all administered tasks, among completed tasks, and among submitted comprehension responses. In the first section, LLM+Search completed seventy-four percent of tasks compared to thirty-one percent for Search-only, and abandoned far fewer tasks, seven percent versus thirty-one percent. Among completed tasks, LLM+Search also had a much higher solution correctness rate, seventy-two percent versus twenty percent for Search-only. Comprehension accuracy, based on a small number of submitted responses, was higher for Search-only, seventy-five percent, than for LLM+Search, fifty-eight percent.}
    \begin{tabular}{lccc}
        \toprule
        & Search-only & LLM+Search & Overall \\
        \midrule

        \multicolumn{4}{l}{\textit{All administered tasks}} \\
        Completed
            & 5/16 (31.2\%)
            & 32/43 (74.4\%)
            & 37/59 (62.7\%) \\
        Timed out
            & 6/16 (37.5\%)
            & 8/43 (18.6\%)
            & 14/59 (23.7\%) \\
        Abandoned
            & 5/16 (31.2\%)
            & 3/43 (7.0\%)
            & 8/59 (13.6\%) \\

        \addlinespace
        \multicolumn{4}{l}{\textit{Among completed tasks}} \\
        Solution correct
            & 1/5 (20.0\%)
            & 23/32 (71.9\%)
            & 24/37 (64.9\%) \\

        \addlinespace
        \multicolumn{4}{l}{\textit{Among submitted comprehension responses}} \\
        Comprehension correct
            & 3/4 (75.0\%)
            & 14/24 (58.3\%)
            & 17/28 (60.7\%) \\

        \bottomrule
    \end{tabular}
\end{table}
}


\newcommand{\TransitionsFigure}{
    \begin{figure}[t]
      \centering
      \includegraphics[width=\textwidth, alt={Six diagrams showing transition patterns among Orienting, Fixing, and Verification states, using one-way and bidirectional arrows.}]{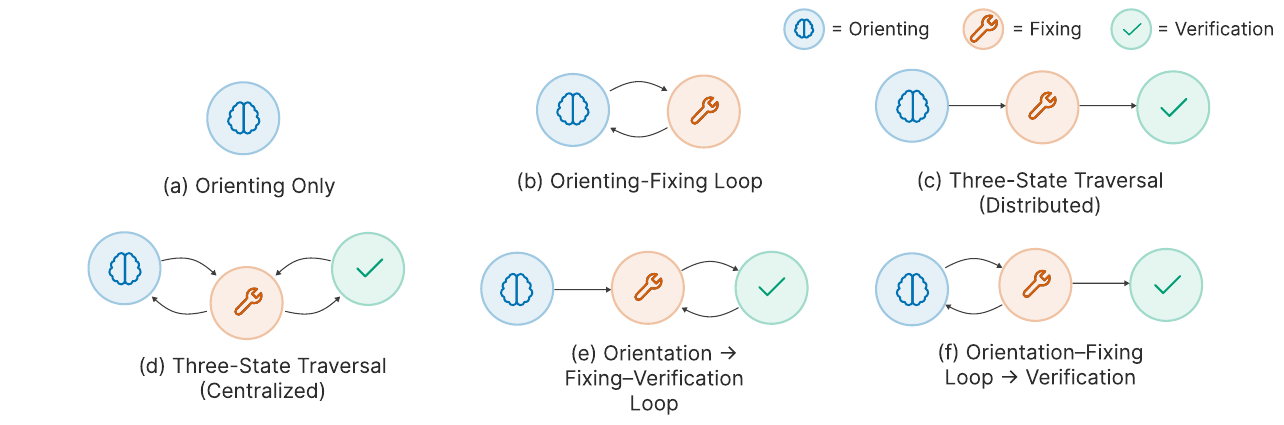}
      \caption{Six observed reasoning patterns characterized by participants’ transitions among orienting, fixing, and verification.}
      \label{fig:state_transitions}
      \Description{Six diagrams depicting transition patterns among the Orienting, Fixing, and Verification states, shown using arrows to indicate direction and looping. Each panel shows one to three state nodes connected by arrows: a single-headed arrow indicates a one-way transition, and two opposing arrows indicate a bidirectional loop between states. Panel a, Orienting Only, shows a single Orienting node with no transitions. Panel b, Orienting-Fixing Loop, shows Orienting and Fixing connected by a bidirectional loop. Panel c, Three-State Traversal, Distributed, shows Orienting, Fixing, and Verification connected by one-way arrows in sequence, moving from Orienting to Fixing to Verification. Panel d, Three-State Traversal, Centralized, shows Fixing connected to both Orienting and Verification by bidirectional loops, with Orienting and Verification not directly connected to each other. Panel e, Orientation to Fixing-Verification Loop, shows a one-way arrow from Orienting to Fixing, followed by a bidirectional loop between Fixing and Verification. Panel f, Orientation-Fixing Loop to Verification, shows a bidirectional loop between Orienting and Fixing, followed by a one-way arrow from Fixing to Verification.}
    \end{figure}
}

\newcommand{\ActivitiesFigure}{
    \begin{figure}[t]
      \centering
      \includegraphics[width=\textwidth, alt={Lists of activities and reasoning behaviors for Query Orienting, Query Fixing, and Query Verification, grouped by state in three columns.}]{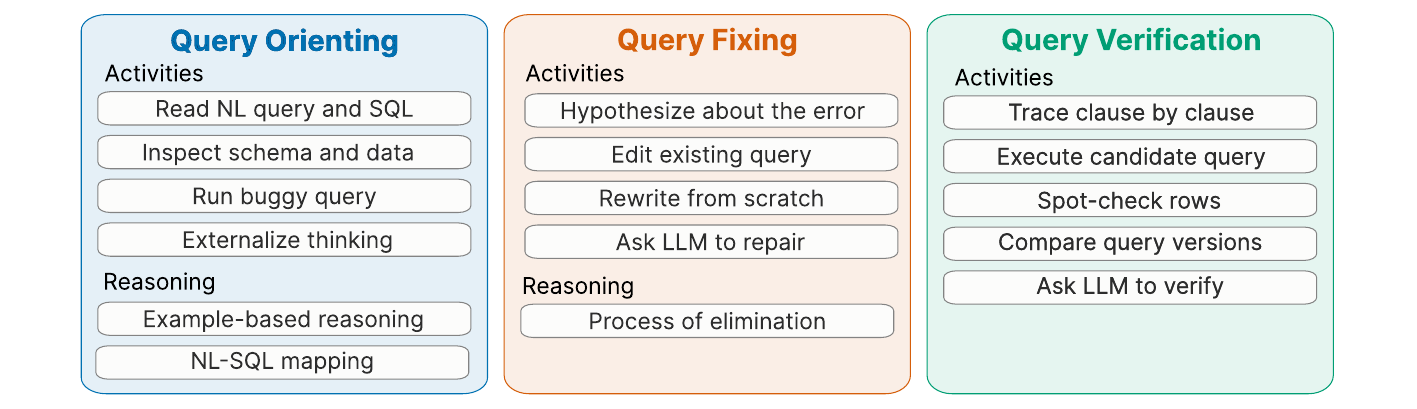}
      \caption{Participant debugging strategies (i.e., basic activities and reasoning behaviors) within each of the Orienting, Fixing, and Verification states}
      \label{fig:activites}
      \Description{Activities and reasoning behaviors listed for the Query Orienting, Query Fixing, and Query Verification states. The Query Orienting column lists four activities: read natural-language query and Structured Query Language, inspect schema and data, run buggy query, and externalize thinking. It lists two reasoning behaviors: example-based reasoning and natural language to structured query language mapping. The Query Fixing column lists four activities: hypothesize about the error, edit existing query, rewrite from scratch, and ask Large Language Model to repair. It lists one reasoning behavior: process of elimination. The Query Verification column lists five activities: trace clause by clause, execute candidate query, spot-check rows, compare query versions, and ask Large Language Model to verify. It lists no reasoning behaviors.}
    \end{figure}
}

\newcommand{\PromptTypeXAutomationExamples}{
    \begin{table*}[t]
    \centering
    \begin{tabular}{p{0.20\textwidth} p{0.74\textwidth}}
    \toprule
    \textbf{Automation level} & \textbf{Representative examples} \\
    \midrule

    \textbf{Low automation}
    & $\bullet$ ``What does this query do? [\textit{buggy SQL}]'' (\textsc{Explain})
    
      \smallskip
      
      $\bullet$ ``Can you elaborate on this business question: [\textit{task requirements}]. Draw me a diagram [\textit{buggy SQL}]'' (\textsc{Generate}) \\

    \addlinespace[5pt]\hline\addlinespace[5pt]

    \textbf{Moderate automation}
    & $\bullet$ ``Is this close? [\textit{candidate solution}]'' (\textsc{Verify})
    
    \smallskip
    
    $\bullet$ ``I don't understand what assumptions this query is making on the data, is there something I need to verify or can I trust this query?'' (\textsc{Strategize}) \\

    \addlinespace[5pt]\hline\addlinespace[5pt]

    \textbf{High automation}
    & $\bullet$ ``I have a SQL query question and a solution, but I don't know why it's wrong. Could you help me find it out? [\textit{task requirements}] [\textit{buggy SQL}]'' (\textsc{Detect})
    
      \smallskip
      
      $\bullet$ ``The goal for this query is [\textit{task requirements}]. I think the issue is that it is still counting bonus points even in months where the customer did not reach 700 units. Can you fix this query?'' (\textsc{Fix}) \\

    \bottomrule
    \end{tabular}
    \caption{Representative examples of LLM interactions at each level of automation.
    Parenthetical labels indicate the prompt type in our taxonomy
    (Appendix~\ref{app:prompt-type}).}
    \label{table:examples}
    \Description{The table has two columns, automation level and representative examples, with three rows corresponding to low, moderate, and high automation. Each row gives two example prompts labeled with their prompt type from the paper's taxonomy: explain and generate prompts for low automation, verify and strategize prompts for moderate automation, and detect and fix prompts for high automation. The examples show requests ranging from asking what a query does, to checking whether a candidate solution is close, to asking the assistant to directly find and fix an error in a query.}
    \end{table*}
}

\newcommand{\AutomationFigure}{
    \begin{figure}[t]
      \centering
      \includegraphics[width=0.9\linewidth, alt={Stacked bar charts: automation differs most by education group, while patterns across Structured Query Language experience are less distinct.}]{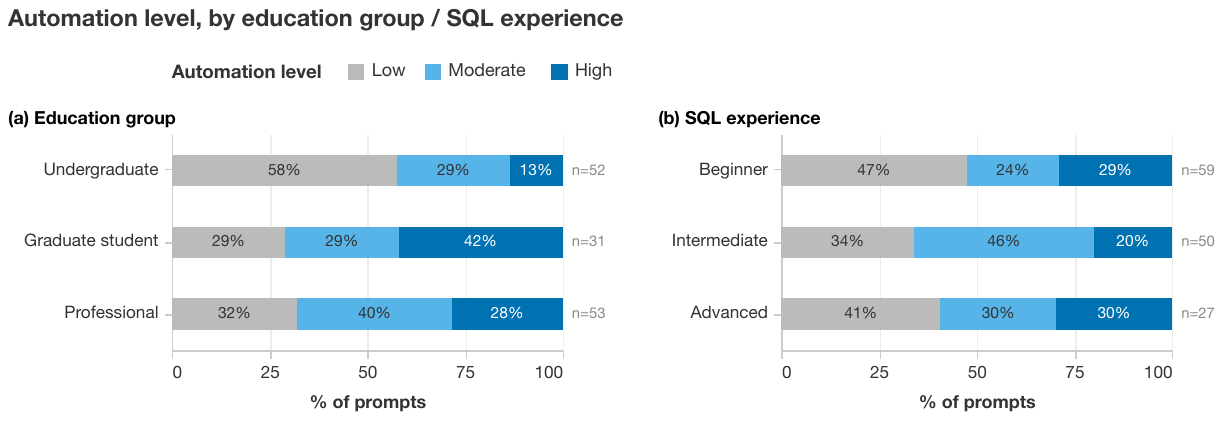}
      \caption{Distribution of prompt automation levels (low/moderate/high) by (a) education group and (b) SQL experience. Bars show the percentage of prompts at each automation level; n indicates the number of prompts.}
      \label{fig:automation}
      \Description{Automation-level distributions by education group and Structured Query Language experience. The education-group panel shows the differences reported in the text. Across Structured Query Language experience levels, beginner prompts were 47 percent low, 24 percent moderate, and 29 percent high automation; intermediate prompts were 34 percent low, 46 percent moderate, and 20 percent high; and advanced prompts were 41 percent low, 30 percent moderate, and 30 percent high. The experience groups contained 59 beginner, 50 intermediate, and 27 advanced prompts; the education groups contained 52 undergraduate, 31 graduate-student, and 53 professional prompts.}
    \end{figure}
}

\newcommand{\ContextFigure}{
    \begin{figure}[t]
        \centering
        \includegraphics[width=\columnwidth, alt={Stacked bar charts: prompts include more task and schema context for graduate, professional, and advanced Structured Query Language users.}]{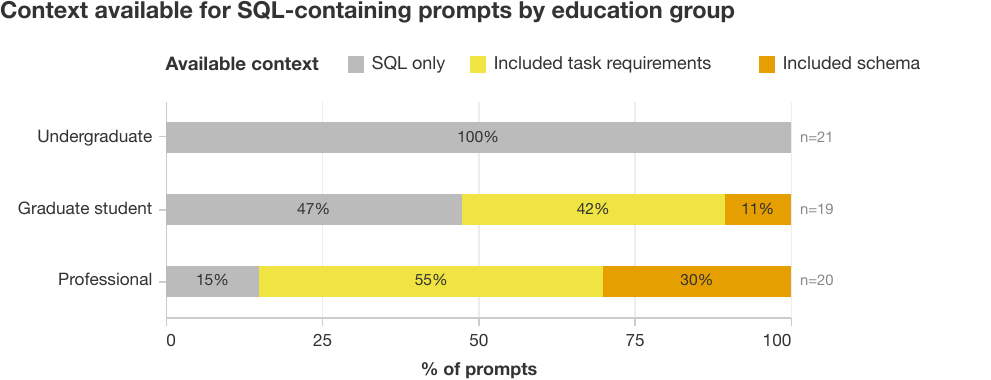}
        \caption{Context available for SQL-containing prompts by education group. Bars show the percentage of prompts in each context category; n indicates the number of SQL-containing prompts.}
        \label{fig:prompt_context}
        \Description{Context provided for SQL-containing prompts by education group. Among undergraduate prompts, 100 percent contained Structured Query Language only. Graduate-student prompts were 47 percent Structured Query Language only, 42 percent with a task description, and 11 percent with schema information; professional prompts were 15 percent, 55 percent, and 30 percent, respectively. Group totals were 21 undergraduate, 19 graduate-student, 20 professional.}
    \end{figure}
}


\newcommand{\PreSurveyFigure}{
    \begin{figure}[t]
      \centering
      \includegraphics[width=\textwidth, alt={Diverging stacked bar chart: trust ratings cluster around 4 to 5, while perceived natural-language-to-Structured-Query-Language accuracy skews high.}]{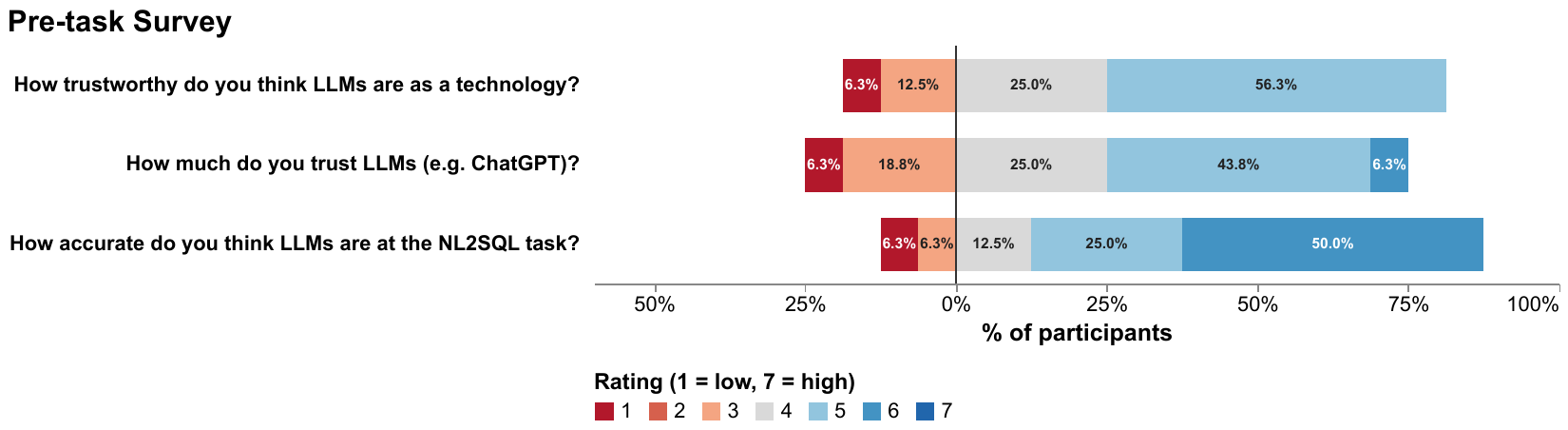}
      \caption{Pre-study survey responses on LLM trust, LLM trustworthiness, and NL2SQL performance}
      \label{fig:presurvey_trust}
      \Description{Pre-task response distributions for trust and perceived natural-language-to-Structured-Query-Language accuracy. Ratings range from 1, low, to 7, high. Trustworthiness and personal trust responses cluster around ratings 4 and 5. Perceived task accuracy is more concentrated at the high end, with half of participants selecting 7, one-quarter selecting 5, and about one-quarter selecting ratings 1 through 3.}
    \end{figure}
}

\newcommand{\NasatlxFigure}{
    \begin{figure}[t]
      \centering
        \includegraphics[width=0.9\textwidth, alt={Diverging stacked bar chart: LLM plus Search shows lower workload and frustration and higher satisfaction than Search-only.}]{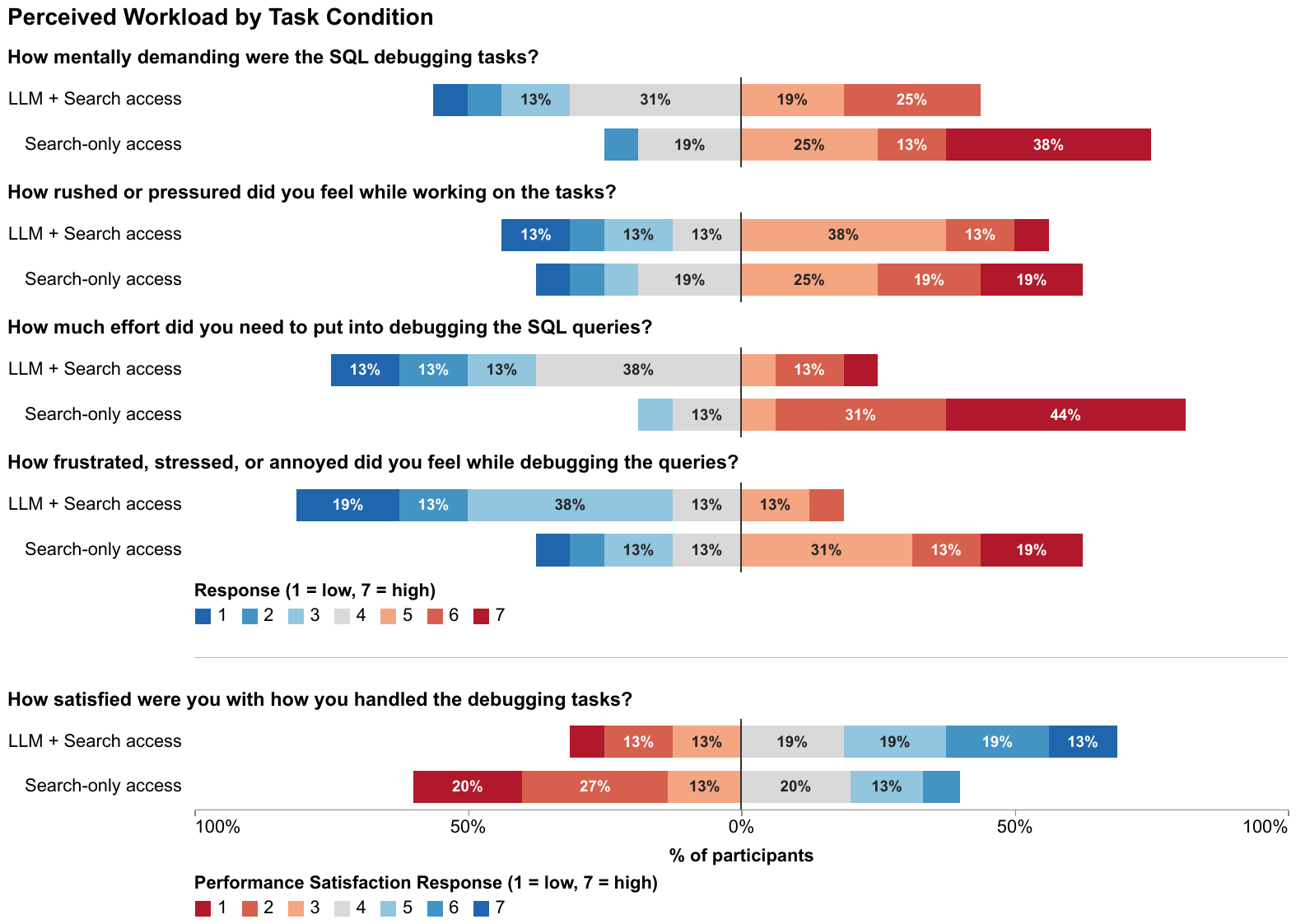}
      \caption{Participant response distributions for NASA-TLX items by study condition: LLM+Search and Search-only.}
      \label{fig:nasatlx}
      \Description{Response distributions for workload and satisfaction across the two access conditions. Ratings use seven-point scales from 1, low, to 7, high. Compared with Search-only access, large language model plus Search responses generally shift toward lower mental demand, time pressure, effort, and frustration, and toward higher satisfaction. The clearest separation is for effort: Search-only responses concentrate toward the high end of the scale, whereas large language model plus Search responses concentrate around the middle and lower end. The other measures show greater overlap between conditions.}
    \end{figure}
}

\newcommand{\LLMResponsesFigure}{
    \begin{figure}[t]
      \centering
        \includegraphics[width=\textwidth, alt={Diverging stacked bar chart of response ratings; satisfaction and trust are highest, followed by understanding and independent confidence.}]{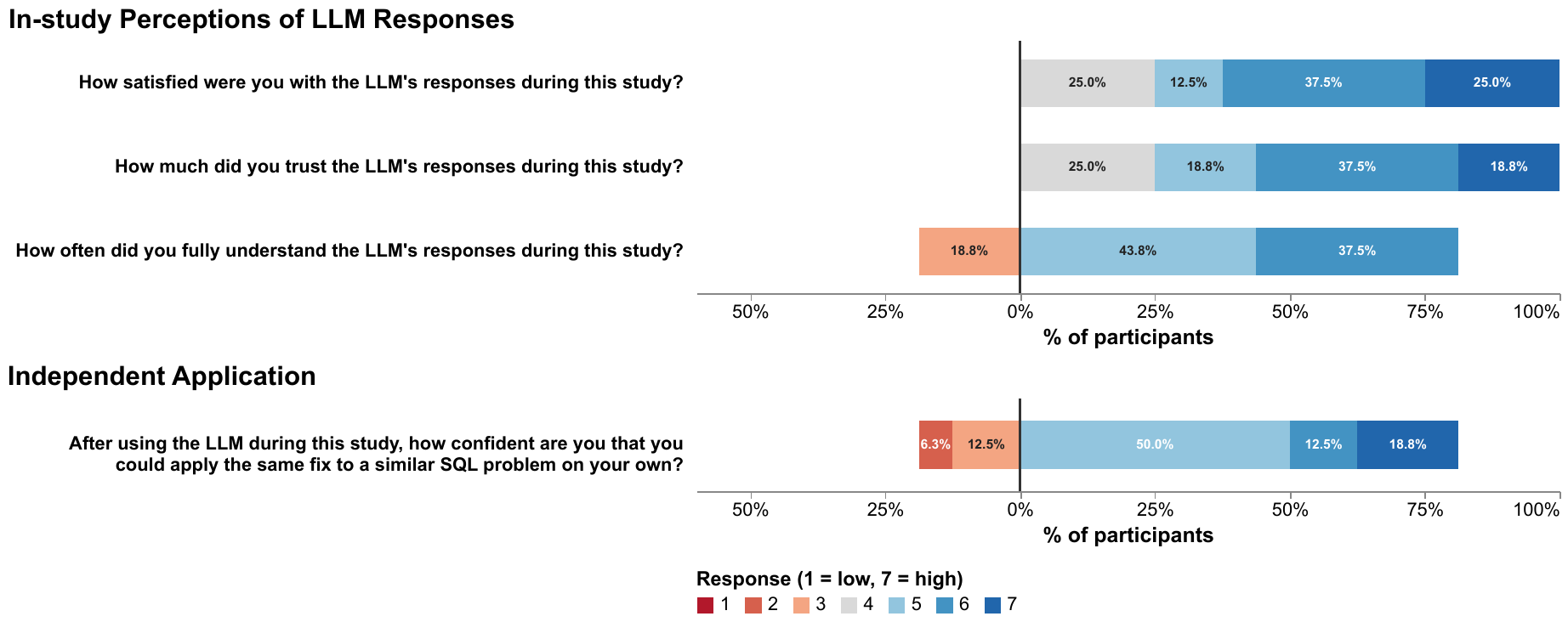}
      \caption{Participant response distributions for in-study perceptions of LLM responses and confidence in applying a similar fix without LLM assistance.}
      \label{fig:llm_responses}
      \Description{In-study response distributions and confidence in independent application. Ratings range from 1, low, to 7, high. Satisfaction and trust responses fall entirely between 4 and 7, with 37.5 percent selecting 6 for each. Understanding responses are concentrated at 5 and 6, totaling 81.3 percent, with 18.8 percent at 3. Confidence in applying a similar Structured Query Language fix independently is also concentrated at 5 through 7, totaling 81.3 percent, while 18.8 percent selected 2 or 3.}
    \end{figure}
}

\section{Introduction}

Large Language Models (LLMs) have become increasingly capable of generating~\cite{DBLP:journals/pvldb/GaoWLSQDZ24, DBLP:journals/tkde/LiuSLMJZFLTL25, DBLP:journals/tkde/LiuZGLLSHGLDZ26, DBLP:conf/nips/PourrezaR23} and debugging~\cite{DBLP:conf/nips/LiLQJQHSHXZTLWZ25, ShenCQZ2026, DBLP:conf/iui/NingZSTZL23} SQL queries with varied amounts of user interaction~\cite{DBLP:conf/chi/ChanSE26, cerebra, dis_db_querying_10.1145/3800645.3812906}. Despite the growing availability of these tools, our fundamental understanding of how users reason through erroneous SQL queries with and without LLM support remains limited. Prior work has categorized common model errors~\cite{DBLP:conf/iui/NingZSTZL23, ShenCQZ2026} and SQL debugging activities~\cite{DBLP:conf/chi/GathaniLB20, Haroon_2024}. However, \emph{why} users pursue certain debugging strategies, \emph{how} LLMs may alter users' SQL reasoning, and \emph{where} current debugging methods fail remain unclear. Furthermore, as users gain access to more capable and more autonomous agents~\cite{chen2026code}, is SQL query debugging becoming a ``solved'' problem from the user’s perspective? And what role might interactive, GUI-based tools play in SQL debugging going forward?

As a first step towards answering these questions, we conducted a user study with 16 participants of varying levels of expertise (8 beginner, 4 intermediate, and 4 advanced SQL users). To test the capabilities of both LLMs and human experts, we focused our study on the task of debugging complex SQL queries, i.e., queries that even experts in database systems would find non-trivial to debug. Participants were asked to complete four debugging tasks: three tasks with access to a general-purpose LLM assistant and one task without LLM assistance. To understand how our participants reason through the SQL debugging process, we qualitatively analyzed transcripts, screen recordings, LLM prompts, SQL queries executed against a provided database, and participant responses to survey questions regarding task difficulty, user confidence, and trust in AI.

Participants' SQL reasoning seemed to remain consistent with observations made in prior studies predating LLMs \cite{DBLP:conf/chi/GathaniLB20, Haroon_2024, DBLP:conf/iui/NarechaniaFLR21}. From these observations, \textbf{we contribute a three-state model of user reasoning during SQL debugging tasks}, which can facilitate modeling of users' SQL debugging strategies in both LLM-assisted and LLM-free debugging tasks.
In our model, participants spend time in three reasoning states: orienting to, fixing, and verifying queries, following varied trajectories across them. We collated our observations to identify common reasoning patterns across participants and tasks, as well as strategic deviations from these patterns facilitated by LLM use.
For example, participants occasionally entered query verification without first developing a repair themselves, instead evaluating a model-generated candidate fix before deciding whether to adopt or modify it. Participants also continued to engage in SQL debugging activities documented in prior work, sometimes using the LLM to support these practices.
Our findings point to \textbf{a small set of recurring mental models} that users rely on as they debug SQL queries, which LLM assistants can complement without disrupting the user's overall reasoning process.

Our findings reveal parallels between SQL query debugging and other coding tasks in terms of how users manage LLM interactions~\cite{chen2026code,zamfirescu-Pereira2025beyond,DBLP:conf/chi/KhuranaSWC25,DBLP:conf/chi/FengYW26, DBLP:conf/chi/MaWZMW25, DBLP:conf/chi/WagmanDC25}. 
Across all levels of expertise, we observed low, medium, and high levels of automation (i.e., how much of the debugging process the user asked the LLM to perform). \textbf{Nearly three quarters of participants' prompts asked the LLM to support their understanding or make bounded progress on the task, rather than directly perform diagnosis or repair}. How participants divided work with the LLM also varied substantially, from asking it to explain a query fragment to asking it to generate the solution, particularly across education groups (undergraduate, graduate student, professional). While participants generally retained control over the debugging process, consistent with prior work on LLM-assisted software development (e.g., \cite{chen2026code,DBLP:conf/chi/KhuranaSWC25}), our findings highlight substantial variation in how users delegated debugging work to the LLM.

Participants seemed to scope their automation preferences based on what role they assigned to the LLM assistant in the debugging process; for example, whether they saw the LLM as more of a calculator, sounding board, or teaching assistant \cite{DBLP:conf/chi/KazemitabaarYWH24, DBLP:conf/chi/IsaWWKTK26}. Participants generally viewed LLM assistance favorably, but their trust was conditioned on what they believed the model could reliably handle and whether it had the appropriate context for the assigned role. Furthermore, \textbf{participants monitored the context they provided to the LLM assistant to ensure that it could not assume a role that they did not want it to have.} Applying these roles allowed participants to seek SQL debugging assistance while retaining responsibility for the solution. In contrast, SQL authoring and debugging tools do not seem to let users manage the full context scope (e.g., \cite{cerebra,DBLP:conf/chi/GathaniLB20,DBLP:conf/uist/TianKLZ24}).

In summary, this paper makes the following contributions:
\begin{itemize}
    \item We contribute a model showing how users move through three states of reasoning as they debug complex SQL queries: \emph{orienting to}, \emph{fixing}, and \emph{verifying} queries. Our findings show how LLMs can augment users' reasoning patterns (i.e., transitions between reasoning states) and complement established debugging practices.
    
    \item We show that LLM-assisted SQL debugging is characterized by selective rather than full delegation, with users primarily seeking support for understanding and bounded progress.

    \item Contrary to prior work on LLM interaction and SQL debugging, our findings show how SQL users withhold context from LLMs to maintain control and enforce specific collaborative roles, such as ensuring the LLM remains a sounding board or teaching assistant throughout the debugging process.
\end{itemize}

\section{Related Work} \label{sec:rel_work}

\subsection{SQL Debugging Strategies and Workflows}

Prior work has characterized debugging strategies for imperative programming~\cite{McCauley_2008, Spörer_Michaeli_2025, Whalley_2023, Aljehane_2021, DBLP:conf/chi/GugertyO86}; SQL's declarative nature introduces distinct debugging challenges~\cite{Almendros_2021}. SQL education work has documented common errors \cite{Smelcer_1995, Taipalus_2018, Taipalus_Perala_2019, Miedema_2021}, but not necessarily how users identify and repair those errors. 
Through interviews with SQL users, Gathani et al. observe a variety of SQL debugging strategies, such as inspecting output statistics, recursive inspection of query fragments, and using trial-and-error to fix bugs~\cite{DBLP:conf/chi/GathaniLB20}.

Others have observed similar debugging behaviors with various debugging tools (e.g., \cite{Haroon_2024,battle2016making}). Narechania et al. designed the DIY debugger to enable users to inspect and modify sample data, and observed users exploring ``what-if scenarios''~\cite{DBLP:conf/iui/NarechaniaFLR21}. These studies document common SQL debugging activities, but overlook how SQL users adjust their debugging strategies over time as they reason through the problem space.

LLM-assisted SQL authoring work has also documented a limited range of SQL debugging activities~\cite{DBLP:conf/chi/ChanSE26,cerebra}
. For example, Zhou et al. observed practitioners moving through prompt formulation, code review, and code refinement while authoring SQL with LLM-driven tools~\cite{cerebra}. 
Here, the core issue is LLM misinterpretation of users' SQL authoring intent. We study the separate problem of users leveraging LLMs to reason through and fix unfamiliar SQL queries.

\subsection{User Perceptions and Use of Human-AI Collaborative SQL Tools}

We observe a rise in LLM-assisted SQL query authoring~\cite{cerebra, DBLP:conf/vl/MiedemaF21}, interpretation~\cite{DBLP:conf/sigmod/LeventidisZDGJR20,DBLP:conf/chi/ChanSE26}, and debugging~\cite{DBLP:conf/iui/NarechaniaFLR21, DBLP:conf/uist/TianKLZ24,DBLP:journals/pvldb/HuCLSZCSRY26,DBLP:journals/pacmmod/HuGSRY24, DBLP:journals/pvldb/HuCLSZCSRY26} tools. We focus on SQL debugging tools, which are designed around a fixed set of debugging mechanisms such as explaining SQL queries step-by-step~\cite{DBLP:conf/iui/NarechaniaFLR21,DBLP:conf/uist/TianKLZ24}, stepwise inspection of intermediate query results~\cite{DBLP:journals/pvldb/HuCLSZCSRY26, Haroon_2024,battle2016making}, or generating minimal edit hints toward a reference query~\cite{DBLP:journals/pacmmod/HuGSRY24}. In contrast, we seek to observe what strategies SQL users adopt in a more open-ended SQL debugging context.

Interestingly, Ning et al. found that advanced SQL users were uninterested in the automated debugging features provided in their user study, and participants distrusted their conversational dialog model~\cite{DBLP:conf/iui/NingZSTZL23}. We test whether this distrust still holds with more powerful LLM assistants and wider adoption of LLMs in data science.

A key challenge in Human-AI collaboration is managing context between the human and model throughout complex tasks~\cite{zamfirescu-pereira2023why,xu2026duetui,DBLP:conf/chi/FengPLASBWZC26}. Prior work on code assistants has found differing preferences over whether systems should automatically use broad context or allow users to control what information is included to the underlying models~\cite{DBLP:conf/chi/McNuttWDD23}. Cerebra addresses missing context as an alignment problem: users may omit schema, domain, or task knowledge needed to generate appropriate SQL, so the system surfaces this information for users to review and refine~\cite{cerebra}. Other work similarly characterizes omission of context as user difficulties in articulating or recalling relevant information~\cite{DBLP:conf/uist/ZhangSXWZQY25}. How users manage context during SQL debugging, however, remains unclear.

\subsection{User Agency and Delegation in Human-AI Collaboration} 

Studies of AI-assisted software distinguish between collaborative and autonomous uses of AI, and show that users often value retaining control over what work is delegated~\cite{DBLP:conf/chi/FengYW26, DBLP:conf/chi/KhuranaSWC25, DBLP:conf/chi/WagmanDC25, chen2026code}. For example, users may prefer copilot-style interaction when they want to understand the underlying work, while others delegate more directly when efficiency is the primary goal~\cite{DBLP:conf/chi/KhuranaSWC25, chen2026code}.
This division of labor can also evolve as planning and execution unfolds~\cite{DBLP:conf/chi/FengPLASBWZC26}.

Beyond task completion, education-focused LLM assistants scaffold tasks without immediately revealing solutions to promote students' independent reasoning and ownership~\cite{DBLP:conf/chi/MaWZMW25, DBLP:conf/chi/KazemitabaarYWH24}. Prior work also reports substantial variation across users and contexts, suggesting that effective human-AI systems should accommodate multiple forms of collaboration rather than assume a single desirable level of automation \cite{DBLP:conf/chi/FengYW26, DBLP:conf/chi/MaKW26, DBLP:conf/chi/WagmanDC25}.

We build on this literature by examining how different divisions of labor manifest during SQL debugging. Instead of asking whether users prefer human or AI control, we study how frequently participants delegate during SQL debugging, what roles they expect the LLM to play, and how they bound its participation in the task.

\section{User Study}\label{sec:user_study}

We conducted a user study examining how participants with varied SQL expertise debug complex NL2SQL outputs, motivated by the following research questions:

\begin{itemize}
    \item[\textbf{RQ1.}] \textbf{Outcomes.} How do task completion, correctness, and comprehension differ between LLM-assisted and unassisted debugging of complex NL2SQL queries?

    \item[\textbf{RQ2.}] \textbf{SQL Debugging Strategies.} What workflows and activities characterize the SQL debugging process, and how does the use of LLMs influence this process?

    \item[\textbf{RQ3.}] \textbf{LLM Interaction.} How do participants choose to interact with the LLM, including what they ask and how they use its responses?
    
    \item[\textbf{RQ4.}] \textbf{LLM Perceptions and Roles.} What are participants' perceptions of LLMs, and how do they relate to their in-study LLM interactions?

\end{itemize}

The study was determined exempt by our university's Institutional Review Board.

\StudyOverview

\subsection{Participants}
We recruited 16 participants (5 F, 11 M) through a mailing list for database practitioners and our professional networks (see Table~\ref{tab:participant-demographics}). We grouped participants by years of SQL experience: less than one year as Beginner (8 total), one to six years as Intermediate (4 total), and more than six years as Advanced (4 total). Participants received a \$25 gift card and a university-branded mug for their participation.

\subsection{Experiment and Task Design}\label{sec:tasks}

\TaskExamples

We used a within-subjects design where 16 participants completed up to four debugging tasks during one 90-minute session (see Figure~\ref{fig:study}): three with access to a general-purpose LLM and search engine (\textit{LLM+Search}) and one with search-engine access only (\textit{Search-only}). The latter served as a baseline for observing debugging without LLM assistance. Task order was counterbalanced using a balanced Latin square, and condition assignment was balanced so that each task served as the Search-only task equally across participants.

In consultation with database researchers at our home institution and through multiple rounds of piloting with knowledgeable SQL users, we developed four complex SQL queries, one per debugging task in our study design (see Figure~\ref{fig:task_examples} for examples). These queries were also informed by existing NL2SQL benchmarks (e.g., \cite{DBLP:conf/nips/LiLQJQHSHXZTLWZ25}) and studies of challenging SQL concepts such as recursion and universal quantification~\cite{DBLP:journals/jeric/TaipalusS20,DBLP:journals/pacmmod/HuGSRY24, DBLP:journals/pvldb/HuCLSZCSRY26, DBLP:conf/sigmod/LeventidisZDGJR20}.

Candidate queries were generated with assistance from Claude Sonnet-4.5 \cite{anthropic2025sonnet45}, which belongs to a different model family than the GPT model used in the study. The authors, including two database experts, reviewed all tasks to ensure the errors were non-trivial and syntactically valid. The full task materials are provided in Appendix~\ref{app:debugging-tasks}.

To approximate a realistic, LLM-assisted debugging experience, participants accessed the LLM through the native ChatGPT interface~\cite{openai_chatgpt_2026} using the default model available (GPT-5.4/5.5; Table~\ref{tab:participant-demographics}), with no restrictions on how they used it. Participants also had access to Google Search with AI-generated summaries disabled and were encouraged to use available tools as they would in their typical debugging workflows.

In our pilots, we found that unassisted tasks took substantially longer to debug, hence their reduction to one per participant. Due to this task imbalance, we treat comparisons between access conditions as exploratory.

\ParticipantInfo

\subsection{Study Procedure and Data Collection}

After providing informed consent, participants were introduced to the study and think-aloud protocol. With permission, we recorded audio and screen activity. Participants were provided with a PostgreSQL database through a terminal, an editable task document, paper and pen, and the tools available in their assigned condition. They were instructed to debug as they would in practice. The interviewer could answer basic PostgreSQL usage questions and clarify task requirements but did not assist with debugging.

Consistent with similar studies~\cite{DBLP:conf/uist/TianKLZ24,cerebra,DBLP:conf/chi/ChanSE26,DBLP:conf/iui/NingZSTZL23}, we observed participants debugging under time constraints.
Participants first completed an LLM-supported warm-up task adapted from a prior NL2SQL study~\cite{DBLP:conf/chi/ChanSE26} (see Appendix~\ref{app:warmup}). The warm-up data were excluded from analysis. Participants then attempted the four main debugging tasks. If a participant did not finish a task within 20 minutes, they moved on to the next task. Participants could also abandon a task. Participants spent 60-70 minutes on the main tasks.

After each task, participants answered questions about their understanding, verification strategy, and perceived difficulty. After the debugging tasks, participants completed a questionnaire assessing workload, perceptions of LLM responses, prior LLM use, and SQL experience. When time permitted, participants also completed a semi-structured interview about their debugging experience and typical use of LLMs. Full survey and interview materials are provided in Appendix~\ref{app:survey-interview-materials}.

Across the 16 sessions, we collected participant transcripts, executed SQL queries and final answers, 22 hours of audio and screen recordings, and 139 LLM prompts.

\section{Data Analysis Methods}\label{sec:analysis_methods}

\subsection{Transcript Analysis}\label{subsec:transcript_analysis_methods} 

We developed the codebook inductively from participant transcripts \cite{DBLP:journals/pacmhci/McDonaldSF19}, beginning with open coding of six transcripts, two per education group (undergraduates, graduate students, professionals), to surface candidate codes. We then iteratively refined the code definitions while coding the remaining 12 transcripts, followed by a final consolidation pass to complete the codebook. The final codebook covered debugging and reasoning strategies, as well as participants' perceptions and evaluations of the LLM (see Appendix~\ref{app:transcript-coding}). We analyzed LLM interactions separately at the prompt level (see Section~\ref{subsec:prompt_analysis_methods}).

Two members of the study team independently coded every transcript. Instead of computing inter-rater reliability (IRR), the coders reconciled all disagreements through discussion. This follows guidance that grounded approaches rarely require IRR and that informal discussions of agreement can indicate consistency of the codes
\cite{DBLP:journals/pacmhci/McDonaldSF19}. 

\subsection{Video Analysis and Pattern Analysis}\label{subsec:video_and_pattern_analysis}

To complement the transcript analysis, the first author coded all video recordings. The codebook was developed from a subset of videos, refined until no new codes emerged, and then applied to the remaining videos (see Appendix~\ref{app:video-coding}). When actions were not verbalized, they were inferred from visible interactions; for example, silently stepping through SQL code was coded as ``SQL Walkthrough.'' 

To characterize broader debugging behaviors, we grouped video codes into three higher-level reasoning states: Query Orientation, Query Fixing, and Query Verification (see Section~\ref{subsec:interaction_patterns}). We used these broader states, rather than finer-grained phases~\cite{DBLP:journals/ijmms/Vessey85}, because participants did not consistently verbalize 
finer-grained reasoning steps.

We represented each task as a transition graph whose nodes were reasoning states and whose edges were observed transitions. The first author clustered these graphs by structural similarity. Ambiguous cases were reviewed against the corresponding video and transition sequence. This analysis yielded six recurring reasoning patterns (see Section~\ref{subsec:interaction_patterns}).

\subsection{Prompt Analysis}\label{subsec:prompt_analysis_methods}

Participants submitted 139 prompts to the LLM. We excluded three accidental submissions, leaving 136 prompts for analysis. We iteratively developed codes along two dimensions: automation and prompt type. Coding focused on the prompt text with session recordings used for further clarification.

\begin{itemize}
    \item \textbf{Automation.} We coded prompts across three levels describing how much debugging work participants delegated to the LLM to perform: \textit{Low} for only supporting participant understanding, \textit{Moderate} for validating a participant-generated approach, and \textit{High} for performing diagnosis, repair, or another part of the debugging process on behalf of the participant.
    
    \item \textbf{Prompt type.} Prompts were grouped by what participants asked the LLM to do, yielding nine request types (\textit{Explain, Detect, Fix, Simplify, Verify, Review, Generate, Strategize, Identify (Syntax)}) and two statement types (\textit{Report, Correction}). Section~\ref{subsec:prompts} analyzes request prompts only.

\end{itemize}

Two general-purpose LLMs (GPT-5.6 Sol and Claude Sonnet-5, both with ``High'' effort) independently evaluated each prompt against the codebook. Following recommendations to validate LLM outputs when models are used as research tools~\cite{DBLP:conf/chi/PangSSBXTB25}, we treated these evaluations as recommendations rather than final coding decisions. For each dimension, models recommended retaining or revising the label, with a brief rationale (see Appendix~\ref{app:llm-coding-review}). The first author reviewed all recommendations and made the final coding decisions based on the prompt and codebook, revising five labels for automation coding and three for prompt type.

For prompts containing or referring to SQL, we coded the task and database context available to the LLM. Two authors jointly defined five context configurations: \textit{SQL Fragment}, \textit{Full SQL}, \textit{Full SQL and Task Requirements}, \textit{Full SQL and Schema}, and \textit{Full SQL, Task Requirements, and Schema}. \textit{Schema} refers to column information from the table. \textit{Task Requirements} refers to text copied directly from the task document, not participants' interpretations. We coded SQL as containing \textit{Schema} or \textit{Task Requirements} if present in the prompt or referenced from earlier chat content.

During video coding, the first author marked LLM-generated code as incorporated when participants visibly integrated all or part of the LLM's suggestions into their work, such as by copying it or referring to the response while modifying their query.

\section{Task Outcomes (RQ1)}\label{sec:results_outcomes}

Here, we summarize participant outcomes and one-shot LLM performance on the same tasks.

\subsection{Task Completion Was Higher with LLM Assistance}\label{sec:results_outcomes:llm_access}

Of 64 planned task attempts, five were omitted due to time constraints. Of the 59 task attempts observed, 16/16 were in the Search-only condition and 43/48 in the LLM+Search condition. Overall, 62.7\% (37/59) of administered tasks were completed, 23.7\% (14/59) timed out, and 13.6\% (8/59) were abandoned. 74.4\% (32/43) of tasks with LLM access were completed versus 31.2\% (5/16) of Search-only tasks (see Table~\ref{tab:task-outcomes}). This result aligns with prior work showing that early, continuous LLM access can improve performance under time pressure over working independently first~\cite{DBLP:conf/chi/ZhiKL26}.

\OutcomesTable

Completion did not always correspond to correctness: 64.9\% (24/37) of completed tasks were correct, and 60.7\% (17/28) of submitted comprehension responses were correct. We do not compare access conditions due to the imbalanced task design. Participants reported a median difficulty rating of 5 on a 7-point scale.

\subsection{LLMs Were Effective at Debugging Most Tasks}\label{rq0_llm_eval}

We evaluated three GPT models with different amounts of prompt context. Each model--prompt--task combination was run 10 times, and responses were evaluated by whether the generated query returned the same rows as the reference query.

\textbf{LLMs achieved high execution accuracy when debugging the provided queries}, with GPT-5.4 reaching 90--100\% accuracy in debug mode, although performance varied across tasks and models. Providing the schema did not consistently improve debugging performance and sometimes reduced accuracy with GPT-5.4. See Appendix~\ref{app:llm-eval} for detailed results by task and model.

\section{Participants' SQL Debugging Strategies (RQ2)}\label{sec:results_activties}

We observed participants moving among three reasoning states while debugging complex SQL queries. In \emph{Query Orienting}, participants interpreted the \nlquery{} and \issuesql{} alongside relevant task components, including the schema and data. In \emph{Query Fixing}, participants worked toward a candidate solution by diagnosing the issue, reasoning about repairs, and editing the query. Prior debugging models separate diagnosis from repair~\cite{DBLP:journals/ijmms/Vessey85}; we combine them because participants did not consistently verbalize diagnosis separately from repair. In \emph{Query Verification}, participants evaluated the correctness of candidate solutions by executing them or tracing their logic.

Participants spent most of their time orienting to the query (median: 5.5 minutes), not fixing (median: 1.3 minutes) or verifying it (median: 1.3 minutes). Similarly, while all attempted task-instances began in Query Orienting, only 92\% reached Query Fixing and 71\% reached Query Verification.

Participants' reasoning did not follow a single Orienting-Fixing-Verification sequence. Instead, they moved through the three states in six recurring patterns, ranging from a linear pass through the states to repeated cycling among them (see Figure~\ref{fig:state_transitions}). These trajectories show how participants applied their own reasoning throughout the debugging process. We characterize these trajectories, then describe the debugging strategies participants used within each state.

\subsection{Debugging Trajectories Reflected Different Sources of Uncertainty}\label{subsec:interaction_patterns}

The six reasoning patterns we observed differed in what participants repeatedly returned to: their understanding of the task, the evolving candidate solution, or both.

\TransitionsFigure

Some trajectories reflected continued uncertainty about the NL task description and the erroneous query. Five instances remained entirely in Query Orienting and never shifted to producing a candidate repair (see Figure~\ref{fig:state_transitions}a). These cases were concentrated in recursive SQL queries: four of five involved recursion, and they ended in both timeouts (3/5) and abandonment (2/5). Another 12 instances (7 recursive, 5 universal-quantifier) reached Query Fixing but not Query Verification (see Figure~\ref{fig:state_transitions}b). Three ended with task completion without verification; the other nine ended because participants ran out of time (5/12) or abandoned the task (4/12). Among these 12 instances, most moved back and forth between Orienting and Fixing as participants revised the query and then returned to understanding the task or query behavior.

Among the 40 instances that reached all three states, participants differed mainly in where they iterated. In 16 instances, no particular state was emphasized: participants either moved through the each state once or cycled through them in sequence (see Figure~\ref{fig:state_transitions}c). In 12 instances, participants returned to one central state (see Figure~\ref{fig:state_transitions}d). Ten were \emph{Fixing-centered}, with participants repeatedly contextualizing, refining, and validating candidate fixes. Two were \emph{Orienting-centered}, with participants repeatedly returning to the task requirements, schema, data, or erroneous SQL. 

Other trajectories showed a clear shift in focus over time. In eight instances, participants oriented to the task, then moved into an iterative Fixing-Verification loop, refining and verifying a candidate without returning to the original task materials (see Figure~\ref{fig:state_transitions}e). In four instances, participants cycled between Orienting and Fixing, but only before verified once at the very end of the task (see Figure~\ref{fig:state_transitions}f). These patterns reveal two sources of uncertainty: lack of clarity on the problem versus the solution.

\textbf{One additional pattern emerged with LLM use, which changed the transition order}. In eight instances, participants entered Verification \emph{before} Fixing because the LLM proposed a candidate solution. Participants evaluated the candidate, then modified or adopted it. This reversed the usual Fixing-before-Verification order. This pattern also allowed participants to retain responsibility for evaluating candidate solutions.

Next, we examine how participants approached the debugging process within each reasoning state.

\ActivitiesFigure

\subsection{Participants Oriented by Connecting Natural Language Intent to Query Behavior}~\label{subsec:QU} 

Before attempting a repair, participants worked to understand how the task requirements mapped onto the \issuesql{} and underlying data. All participants inspected the task, query, schema, or table data, and most (12/16) executed the erroneous query to observe its behavior. Nine participants also modified the query to inspect intermediate results or isolate the behavior of particular query fragments (i.e., clauses or subqueries).

\textbf{Participants used two recurring strategies to reason about query semantics.} Example-based reasoning, seen in 13/16 participants, involved constructing concrete input scenarios and tracing what the \issuesql{} would return for each scenario. These examples helped participants make abstract query behavior more concrete and identify cases in which the output would violate the task requirements. NL-SQL mapping, seen in 12/16 participants, involved systematically relating elements of the natural-language \nlquery{} to corresponding SQL fragments. This helped participants identify which parts of the query implemented each requirement.

One participant (P3) extended example-based reasoning by constructing a small synthetic dataset and executing the \issuesql{} against it, making the semantic discrepancy easier to observe. Together, these strategies show that orienting involved making SQL semantics concrete by connecting intent, query structure, and data behavior.

\subsection{Participants Repaired Queries Incrementally Rather Than Rewriting from Scratch}\label{subsec:QF}

Where Orienting isolates which task requirements are being violated by the \issuesql{}, Query Fixing focuses more on detecting erroneous query fragments and developing candidate repairs. We analyze these activities together because participants often oscillated between diagnosing the error and reasoning about how to fix it, rather than treating them as separate steps.

\textbf{Participants sometimes used process of elimination to detect errors.}
We observed several instances in which participants examined individual SQL clauses or subqueries, ruled out fragments that appeared consistent with the task requirements, and narrowed their attention to the remaining possible sources of error. Once they identified a possible source of the error, \textbf{participants produced candidate solutions by modifying the existing query, requesting an LLM-generated repair, or rewriting the query from scratch}. However, rewriting from scratch was rare (8/59 task instances).

\subsection{Participants Verified Queries by Tracing Logic and Inspecting Outputs}~\label{subsec:QV} 

In Query Verification, participants evaluated whether a candidate solution met the task requirements. Most participants (13/16) traced the candidate SQL clause by clause, assessing whether each fragment aligned with the intended behavior. Similarly, 13/16 participants executed candidate queries to check for runtime errors and inspect whether the resulting output appeared correct.

\textbf{Participants also used targeted checks to test query behavior akin to unit testing.} Five participants ran exploratory versions of candidate queries to inspect intermediate results or isolate the effects of particular clauses. Three participants spot-checked individual output rows against the task requirements, while others compared the outputs of the \issuesql{} and candidate query, or of multiple candidate queries. These comparisons examined differences in returned rows or aggregate indicators such as row counts.

Some participants also considered verification through controlled test data. P08 and P15 reported that they would construct small databases to test candidate queries if time permitted. P12 similarly asked the LLM to generate diagnostic test cases designed to expose incorrect query behavior. These practices overlap with strategies reported in prior SQL debugging work, including inspecting data and query outputs, examining intermediate results or query components, and using small or manipulated datasets to test query behavior \cite{DBLP:conf/chi/GathaniLB20, Haroon_2024, DBLP:conf/iui/NarechaniaFLR21,battle2016making}.

\paragraph{\textbf{Takeaways}} We presented three states of reasoning that our participants moved through as they debugged complex SQL queries, and showed how these reasoning states influenced their selected debugging activities. Many debugging practices described in prior SQL work remained present when participants utilized LLM assistance. For example, participants still inspected data and intermediate results, traced query logic, compared outputs, and reasoned through concrete examples~\cite{DBLP:conf/chi/GathaniLB20,Haroon_2024}. In many cases, the LLM supported these practices, for example, by generating diagnostic test cases or helping participants reason through query logic. In other cases, it changed how debugging unfolded by supplying a candidate solution before participants had developed a repair themselves. Overall, participants incorporated LLM assistance into familiar debugging practices rather than replacing those practices altogether.

\section{LLM Interactions During SQL Debugging (RQ3)}\label{sec:results_llm}

We examined how participants interacted with the LLM during debugging, focusing on how they divided debugging work, formulated requests, and managed the context available to the model. Across the 43 tasks where the LLM was available, participants sent 139 prompts (mean=3.23 per task), of which 136 were analyzed after excluding accidental prompts. On average, participants spent 4.4 minutes interacting with the LLM (30.2\% of task time). Participants varied significantly in how they incorporated the LLM into their debugging process.

\PromptTypeXAutomationExamples

\subsection{Participants Varied in How They Divided Debugging Work with the LLM}\label{subsec:prompts} 

First, we analyze how much debugging work participants wanted to automate using the LLM. Prior work defines automation in terms of the level of decision-making handled by an intervention versus the user~\cite{DBLP:conf/iui/DongBPSBW25}. We identified three levels of automation used by participants:

\begin{itemize}
\item \textbf{Low Automation} (\textit{Help Me Understand}): The LLM helps the participant develop their own understanding of the SQL or task and does not directly advance the debugging process.
\item \textbf{Moderate Automation} (\textit{Help Me Progress}): The LLM performs a limited step that advances the debugging process without taking over the diagnosis or repair. 
\item \textbf{High Automation} (\textit{Do It For Me}): The LLM is asked to detect the error, fix the query, or revise the solution. The LLM's work covers part or all of the current reasoning state (see Section~\ref{sec:results_activties}).
\end{itemize}

\textbf{Many prompts kept the LLM in a supporting role rather than asking it to debug for the participant.} Low automation was most common (41.2\% of prompts), followed by moderate (33.1\%) and high automation (25.7\%). This variation in LLM-use is also reflected in prior work~\cite{chen2026code,DBLP:conf/chi/FengYW26,DBLP:conf/chi/KhuranaSWC25,DBLP:conf/chi/WagmanDC25}.

Participant requests reflected these differences in automation (see Table~\ref{table:examples}). Low-automation prompts primarily supported participants in building the understanding they needed to continue debugging on their own. For example, P03 explicitly constrained the LLM's assistance while asking for an explanation: ``what does this query do? and do not change anything about it.'' Some low-automation prompts also involved generation requests for supporting artifacts, such as examples or formal representations, without directly correcting the SQL. Moderate-automation prompts asked the model to help with a bounded debugging step, such as reviewing a hypothesis, identifying useful SQL syntax, strategizing next steps, or verifying a candidate query. For example, P14 asked, ``Is the bug that the g.start\_consumption is always comparing to the original/base case month instead of the prior month?'' High-automation prompts delegated diagnosis, repair, or query simplification such as P11's request, ``Can you give the non recursive solution?'' The full distribution of request types is provided in Appendix~\ref{app:prompt-type}.

\textbf{The clearest differences in automation appeared across education groups} (see Figure~\ref{fig:automation}a). Undergraduates' prompts were most often low automation (57.7\%), compared with professionals (32.1\%) and graduate students (29.0\%). Professionals had the highest proportion of moderate-automation prompts (39.6\%), while graduate students had the highest proportion of high-automation prompts (42.0\%). Overall, undergraduates more often used the LLM to support understanding, professionals more often used it to advance an existing approach, and graduate students more often delegated diagnosis or repair. In contrast, automation patterns were less distinct across SQL experience levels (see Figure~\ref{fig:automation}b), suggesting that education group better explained differences in LLM use. This observation contrasts with prior findings that more technically experienced students request explanations more often than novice students~\cite{DBLP:conf/chi/MaKW26}: in our study, undergraduates used low-automation prompts, including explanations, more often than graduate students.

\AutomationFigure

Participants incorporated model-generated code in 29 of the 43 tasks where the LLM was available (67.4\%). Incorporation occurred in 35.7\% of undergraduate task attempts, compared with 92.9\% of graduate-student and 73.3\% of professional task attempts. In contrast, incorporation varied relatively little across SQL-experience groups. Prior work has found that directly adopting LLM-generated solutions can undermine learners' sense of achievement and leave them feeling like they are cheating, which may help explain the lower incorporation rate we observed among undergraduates~\cite{DBLP:conf/chi/MaWZMW25}. Together, these patterns suggest that participants differed not only in what they asked the LLM to do, but also in how frequently they integrated LLM outputs into the final solution.

\subsection{Participants Limited the Context Available to the LLM}\label{subsec:context}

Participants also varied substantially in how much of the task and database context they provided to the LLM. For the 60 prompts containing or referring to SQL, we coded whether relevant task information (i.e., task requirements and schema) was available either in the prompt itself or earlier in the chat history. In most cases, participants provided only part of the context available to them. In 33 prompts (55.0\%), participants included SQL without additional task or database information. Another 19 prompts (31.7\%) included the task requirements, while only eight (13.3\%) included schema information. Of these eight prompts, five included the full query, task requirements, and schema, giving the LLM the full context available to the participant.

\ContextFigure

Prompt context also differed by education group (see Figure~\ref{fig:prompt_context}a). Undergraduates never included a task description or schema in their SQL-containing prompts: they provided either a SQL fragment or the full query alone. In contrast, task descriptions appeared in 42.1\% of graduate students' prompts and 55.0\% of professionals' prompts. Professionals also included schema information most often (30.0\%), compared with 10.5\% for graduate students and none for undergraduates. This pattern is consistent with prior observations that professional NL2SQL users frequently supplement queries with schema and business context to bridge knowledge gaps~\cite{cerebra}.

In some cases, \textbf{participants appeared to withhold information from the LLM to constrain the kind of assistance it could provide}. We identified seven prompts that explicitly asked the model to avoid giving an answer, five of which came from undergraduates. For example, P01 asked, ``does this query do something wrong? just a small hint'' while providing only a SQL fragment. Providing less task or schema context, or providing only fragments of SQL, may have similarly allowed participants to seek targeted assistance without giving the model enough information to solve the entire task, curating the kind of help they wanted. However, context omissions may also have been unintentional, particularly among less experienced users who may not have recognized which information was needed. Such omissions resemble those reported in prior work, which have been attributed to LLM inexperience~\cite{DBLP:conf/chi/MaKW26} or delayed articulation or limited memory~\cite{DBLP:conf/uist/ZhangSXWZQY25} rather than deliberate withholding.

\paragraph{\textbf{Takeaways}} 
Participants differed in how much debugging work they asked the LLM to perform, with the clearest differences appearing across education groups (i.e., undergraduate, graduate, professional) rather than SQL expertise. Participants also differed in how much context they provided to the model, and some participants even appeared to constrain their requests and/or withhold information from the LLM. These behavioral patterns describe how participants bounded the LLM's participation in debugging. In the next section, we examine participants' perceptions of the LLM and how the roles they assigned to it help explain these interaction choices.

\section{Participant Perceptions of LLMs (RQ4)}\label{sec:results_perceptions}

In this analysis, we examine how participants perceived the LLM and how these perceptions shaped their interaction choices.

\subsection{Participants Viewed LLMs Favorably but Calibrated Their Trust}\label{subsec:all_perceptions}

Participants generally entered the study with favorable perceptions of LLMs. In the pre-task survey, median Likert ratings were above the midpoint for LLM trustworthiness (Mdn=5.0), personal trust (Mdn=4.5), and perceived NL2SQL accuracy (Mdn=5.5) (see Figure~\ref{fig:presurvey_trust}).
Most participants were also frequent LLM users: 75\% reported frequent or very frequent use in the previous week.

After completing the tasks, participants reported lower mental demand (Mdn=4.0 vs. 5.5), effort (Mdn=4.0 vs. 6.0), and frustration (Mdn=3.0 vs. 5.0) in the LLM+Search condition than in Search-only, with greater performance satisfaction (Mdn=4.5 vs. 3.0) (see Figure~\ref{fig:nasatlx}). Participants also evaluated the in-study LLM responses favorably: satisfaction and trust received median ratings of 6, while reported understanding and confidence in applying a similar fix independently received median ratings of 5 (see Figure~\ref{fig:llm_responses}). Full survey distributions are provided in Appendix~\ref{app:post-survey}. 

\PreSurveyFigure

\NasatlxFigure

\LLMResponsesFigure

Despite these favorable perceptions, participants' trust in the LLM was conditional. Some were more willing to trust the model for tasks they perceived as manageable. P04 noted that ``with simpler queries GPT... doesn't tend to make mistakes.'' Similarly, P08 constrained requests to tasks they expected the model could handle reliably: ``I don't expect [it] to just give me the full answer... I just give it manageable tasks.'' Prior experience also shaped these expectations; for example, P15 reported using the ``LLM without doubt'' for syntax-related requests after repeatedly observing good performance on such tasks. These accounts suggest that participants calibrated trust based on what they believed the LLM could reliably handle, consistent with prior observations that users become more cautious as task complexity increases~\cite{DBLP:conf/uist/ZhuCLZSWW25}.

Trust also depended on whether participants believed the LLM had appropriate context for the request. P07 realized that the model ``has no idea what the schema looks like,'' which changed how they prompted it moving forward. P09 similarly reported greater trust in workplace LLMs because ``they actually have context on our data.'' However, more context was not always perceived as beneficial. P11 deliberately withheld the task requirements because they might ``influence how it interprets the query.'' This demonstrates how participants' beliefs about context shaped their prompting and how much they trusted the LLM's responses. These accounts help explain the context-management behaviors observed in Section~\ref{sec:results_llm}: participants adjusted what information was appropriate for the assistance they wanted.

\subsection{Participants Assigned the LLM Roles that Bounded Its Participation}\label{subsec:roles}

Participants described interacting with the LLM in different roles, particularly as a teaching assistant (TA) or sounding board rather than solely as a calculator that produced solutions. These roles help explain the variation observed in Section~\ref{sec:results_llm}: TA-like use often kept the LLM in a supporting role, sounding board use involved bounded assistance or review, while calculator-like use more directly delegated debugging work.

\paragraph{\textbf{LLM as a TA}}

Several participants described using the LLM to support understanding rather than obtain an answer directly. P04, an undergraduate student, described receiving guidance from instructors to use LLMs to ``help you explain concepts,'' characterizing the LLM as a ``de facto TA.'' This framing was reflected in participants' prompting behavior. For example, P05 anticipated that ChatGPT might provide the solution and instead asked it to ``break this down'' but ``don't tell me yet,'' seeking help identifying ``where the critical problem lies'' before seeing the answer. 

For some participants, limiting the LLM's role was explicitly connected to learning. P04 explained, ``I think I learn a lot less'' when relying on ChatGPT too early in the process. P02 shared a similar sentiment: ``if I just look at the answer, I'm like, oh, okay, I know how to do this. But if I wake up tomorrow... I'm like, wait, what did I do?''. These concerns echo findings that easy access to complete LLM-generated solutions can leave learners' algorithmic thinking underdeveloped~\cite{DBLP:conf/chi/MaWZMW25,DBLP:conf/chi/KazemitabaarYWH24}. By assigning the LLM to a TA role, participants sought assistance while preserving opportunities to reason through the problem themselves.

\paragraph{\textbf{LLM as a Sounding Board}}

Other participants described using the LLM as a reasoning partner or reviewer. P08 explained, ``I try to use ChatGPT to help me think'' because ``[i]f it's just giving a solution and telling me it's correct, I cannot verify it.'' P13 similarly described using the LLM as a ``reviewer'' or ``someone who points out my mistakes,'' rather than trying to solve the problem with it. P15 valued simply ``having a second eye.'' Across these accounts, the LLM supported reasoning and review while participants retained responsibility for judging the correctness of the solution.

This division of responsibility also allowed participants to maintain ownership of the task. P11 emphasized, ``I don't want to just completely devolve responsibility,'' while P14 described using ChatGPT to ``fact [check] my logic'' because they preferred to develop solutions themselves. These accounts resemble prior observations that experienced developers maintain agency through bounded or detailed delegation rather than full reliance~\cite{DBLP:conf/chi/FengYW26}. Together, this demonstrates that participants considered whether the role of the LLM supported their goals for learning, reasoning, and responsibility.

\paragraph{\textbf{LLM as a Calculator}}

Consistent with prior work~\cite{DBLP:conf/chi/WagmanDC25, DBLP:conf/chi/KhuranaSWC25, DBLP:conf/chi/IsaWWKTK26}, some participants delegated substantial portions of the task to the LLM. P06 described initially ``plug and chugging'' by providing the \nlquery{} and buggy query, and asking for the correct solution. This direct delegation may have reflected task difficulty, time pressure, or broader norms around LLM use. For example, P12, a participant in industry, explained ``So it's 2026... what I would probably do, because I'm busy, is just paste this query into my LLM and let it tell me what's wrong.''

\paragraph{\textbf{Takeaways}} Participants generally trusted and valued LLM assistance, but calibrated that trust according to the task, the context available to the model, and the role they wanted it to play. These perceptions help explain the behavioral variation in Section~\ref{sec:results_llm}: choices about what to delegate and what context to provide were tied to whether participants wanted the LLM to teach, review, or more directly perform debugging work.
\section{Discussion}\label{sec:discussion}

In this paper, we presented a controlled user study with 16 participants of varying SQL expertise debugging complex SQL queries with and without the assistance of GPT-5.4/5.5 (see Section~\ref{sec:user_study}). We qualitatively analyzed participant transcripts, screen captures, LLM prompts, SQL queries executed against a provided database, and survey responses (see Section~\ref{sec:analysis_methods}). Overall, our findings show how LLMs are already capable SQL debuggers on their own (see Section~\ref{sec:results_outcomes}), and when assisting human debuggers (see Section~\ref{sec:results_llm}), can augment user reasoning to ease the process of debugging complex SQL queries (see Section~ \ref{sec:results_activties}). However, our results also showed how users carefully managed task context and even withheld information to ensure that LLM assistants remained in the roles that participants assigned to them (see Section~\ref{sec:results_llm}). Otherwise, participants could lose trust in the LLM's suggestions or hinder their ability to learn during debugging tasks (see Section~\ref{sec:results_perceptions}). In this section, we review our findings and their implications for future research in task automation and LLM-assisted SQL debugging.

\subsection{Findings and Implications}
\label{sec:discussion:findings-implications}

\subsubsection{Modeling User Reasoning}
Our findings revealed consistent patterns in how participants reasoned through erroneous SQL queries both with and without LLM assistance, which we synthesized into a three-state model of user reasoning during SQL debugging tasks. Furthermore, we observed how LLM usage altered state transition patterns, showing how LLM assistance can augment users' SQL reasoning patterns. Our work extends prior observations of SQL debugging strategies~\cite{DBLP:conf/chi/GathaniLB20,DBLP:journals/pacmmod/HuGSRY24,DBLP:journals/pvldb/HuCLSZCSRY26,cerebra,ShenCQZ2026,DBLP:conf/iui/NingZSTZL23}, e.g., by considering more complex SQL queries, comparing the strategies of novices and experts, and observing the debugging capabilities of current LLMs. As another example, our findings show how participants' reasoning strategies are not necessarily predicted by SQL expertise and are not always indicative of task performance (see Sections~\ref{sec:results_activties} and \ref{sec:results_llm}). Thus, SQL debugging tools should accommodate variance in reasoning strategies to support a wide range of SQL users.

\subsubsection{Debugging Performance}
The GPT models that we tested performed well on fully automated debugging tasks (see Section~\ref{rq0_llm_eval}). 
Paired with the time savings that they provided during our user study (see Section~\ref{sec:results_outcomes:llm_access}), our findings suggest that new tools will need to be strategic in how they enhance the user's SQL debugging experience beyond what general purpose LLM assistants already provide. One potential improvement is to \textbf{expand context management capabilities for SQL users}. In particular, we do not observe much support for withholding context from LLMs to preserve user agency during SQL debugging tasks (e.g.,~\cite{DBLP:conf/uist/TianKLZ24,cerebra}). Future tools can help users fully externalize the context to be managed (the natural-language prompt, the underlying dataset and schema, the current SQL query) and ensure that key information remains hidden from the LLM, such as withholding the original natural-language prompt or the erroneous SQL query to prevent the LLM from switching to solution generation. Future SQL debugging tools could also establish preset roles for the LLM that align with common user intents, such as allowing users to specify that the LLM should act as a TA or sounding board.

\subsubsection{Beyond SQL} Our findings connect SQL debugging to related research on LLMs as coding assistants~\cite{chen2026code,DBLP:conf/chi/KhuranaSWC25,zamfirescu-Pereira2025beyond,DBLP:conf/chi/FengYW26} and data science assistants~\cite{DBLP:conf/chi/WagmanDC25,shen2026debugging,DBLP:conf/chi/McNuttWDD23}, revealing cross-area overlaps in preferences for user agency and LLM automation. Thus, our research can also help database researchers orient themselves within the broader landscape of research in HCAI and LLM coding assistants.

\subsection{Limitations}
\label{sec:discussion:limtations}

We opted to study SQL queries of greater complexity compared to prior work (e.g., \cite{cerebra,DBLP:conf/uist/TianKLZ24,DBLP:conf/chi/ChanSE26,DBLP:conf/iui/NingZSTZL23}) to test the debugging capabilities of current LLMs and SQL users. As a consequence, our selected queries take more time to debug compared to those observed in prior work, limiting the number of tasks that could be observed in our study. Future work could extend our study methods to include additional queries, or even vary the queries by complexity to understand how user reasoning might change by query complexity.

Our pilot study revealed that the Search-only condition took longer than the LLM+Search. Consequently, we chose an unbalanced design with three LLM+Search tasks and one Search-only for increased opportunities to observe completed debugging sessions. As a result, condition comparison should be treated as exploratory. To target the completion of four tasks, participants were limited to 20 minutes each, consistent with the use of time-limited SQL reasoning tasks in prior work~\cite{DBLP:conf/uist/TianKLZ24,cerebra,DBLP:conf/chi/ChanSE26}. Given evidence that outcomes of LLM use can vary with time availability \cite{DBLP:conf/chi/ZhiKL26}, this constraint may have shaped the LLM use we observed. Future work could include longer sessions or field studies observing SQL debugging in longer time contexts.

We used the default ChatGPT model, which upgraded from GPT-5.4 to GPT-5.5 for the final three participants. This version change may have affected participant behavior for these participants, and reflects broader reproducibility concerns associated with studying evolving proprietary LLMs~\cite{DBLP:conf/chi/PangSSBXTB25}. Future work can either keep the model version fixed for all participants, or purposefully examine how different model versions and families affect debugging behaviors.

\section{Conclusion}\label{sec:conclusion}

A think-aloud study with 16 participants revealed that, when faced with complex SQL debugging tasks, trust in the LLM was high, yet a majority of participants used it to help them understand or incrementally progress rather than simply doing the task for them. These differing levels of automation were characterized by educational group, with graduate students using LLMs in high-automation contexts more often than undergraduates and professionals. Participants also assigned LLMs distinct roles, such as a TA or sounding board, with undergraduates explicitly describing these roles in relation to guidance from their instructors. To keep LLMs within these roles, participants actively controlled the context they provided, withholding information to prevent unintended high-automation actions or providing more context when they wanted a better result. We also contribute a three-state model of user reasoning during SQL debugging tasks, which shows how core SQL debugging strategies do not disappear in the presence of LLMs. However, LLM use sometimes changed participants' debugging paths, such as by bringing query verification before fixing. These findings suggest that LLM-mediated debugging is not just a shift toward greater automation; it is an active process in which users decide to what extent the LLM participates in the workflow. Our findings suggest new directions for designing LLM-assisted SQL debugging tools, such as offering preset role assignments and more comprehensive context management to preserve user agency throughout the debugging process.

\end{sloppypar}







\begin{acks}
This material is based upon work supported by the National Science Foundation Graduate Research Fellowship Program under Grant No. DGE-2140004, and by the National Science Foundation under Grant Nos. IIS-2211133, IIS-2402718. Any opinions,
findings, and conclusions or recommendations expressed in this material are those of the author(s) and do not necessarily reflect the views of the National Science Foundation.

\end{acks}

\section*{Ethics and Privacy Statement}

The study was evaluated by our university's Institutional Review Board and determined to be exempt due to the low risk of harm for participants. Participants were briefed and gave informed consent prior to participating in the study. Participants were not subjected to additional risks beyond what they might normally encounter while using GPT models outside of the study. We believe this work poses minimal risk of societal harm and will not lead to serious privacy, safety or fairness concerns. We believe our research follows the ACM's policies and our university's policies for ethical research.


\bibliographystyle{ACM-Reference-Format}
\bibliography{ref}

\appendix

\section{Task Materials}
\label{app:study-materials}

\subsection{SQL Debugging Tasks}
\label{app:debugging-tasks}

Participants completed four SQL debugging tasks. Tasks A and B involved universal quantification, while Tasks C and D involved recursive queries. Each task presented participants with natural-language task requirements and a syntactically valid SQL query containing a semantic error.

\subsubsection{Task A: Universal Quantification}
\label{app:task-a}

\paragraph{Task Requirements.}
Your analytics team wants to identify customers for a loyalty campaign: customers who have only ever visited gas stations where the station's official product catalog includes their preferred product. Every station on their visit record must stock what they prefer.

\paragraph{Buggy SQL Query.}

\begin{lstlisting}
SELECT DISTINCT v.customerid
FROM visits v
WHERE EXISTS (
  SELECT 1
  FROM station_catalog sc
  JOIN favors f
    ON f.customerid = v.customerid
   AND f.productid = sc.productid
  WHERE sc.gasstationid = v.gasstationid
);

\end{lstlisting}

\subsubsection{Task B: Universal Quantification}
\label{app:task-b}

\paragraph{Task Requirements.}
Your procurement team wants to find customers with a fully contained purchase profile: customers who have only ever visited gas stations where every product in that station's official catalog is something they've personally purchased before. No station on their visit record should carry even a single product outside their purchase history.

\paragraph{Buggy SQL Query.}
\begin{lstlisting}
SELECT v.customerid
FROM visits v
WHERE NOT EXISTS (
  SELECT 1
  FROM station_catalog sc
  WHERE sc.gasstationid = v.gasstationid
    AND EXISTS (
      SELECT 1
      FROM purchases p
      WHERE p.customerid = v.customerid
        AND p.productid = sc.productid
    )
);
\end{lstlisting}

\subsubsection{Task C: Recursive Query}
\label{app:task-c}

\paragraph{Task Requirements.}
The growth analytics team wants to identify customers who have shown sustained consumption growth over at least 3 consecutive calendar months at any point in their history. These customers are candidates for a top-tier engagement program.

\paragraph{Buggy SQL Query.}
\begin{lstlisting}
WITH RECURSIVE growth AS (
    SELECT customerid,
           date::INT        AS cur_month,
           consumption      AS start_consumption,
           1                AS len
    FROM yearmonth

    UNION ALL

    SELECT ym.customerid,
           ym.date::INT,
           g.start_consumption,
           g.len + 1
    FROM yearmonth ym
    JOIN growth g
      ON  ym.customerid = g.customerid
      AND ym.date::INT  = CASE
                            WHEN g.cur_month % 100 = 12
                            THEN (g.cur_month / 100 + 1) * 100 + 1
                            ELSE g.cur_month + 1
                          END
      AND ym.consumption > g.start_consumption
)
SELECT DISTINCT customerid
FROM growth
WHERE len >= 3;
\end{lstlisting}

\subsubsection{Task D: Recursive Query}
\label{app:task-d}

\paragraph{Task Requirements.}
Customer bonus points progress over consecutive months based on usage. For any month where usage is at least 700 units, only the amount above 700 contributes to the customer’s bonus points total. In months that do not meet this threshold, previously accumulated points do not carry forward. Determine the first month in which the customer’s points exceeds 1000.

\paragraph{Buggy SQL Query.}
\begin{lstlisting}
WITH RECURSIVE loyalty AS (
    SELECT
        customerid,
        date::INT AS date,
        GREATEST(consumption - 700, 0) AS running_credit
    FROM yearmonth
    WHERE (customerid, date) IN (
        SELECT customerid, MIN(date)
        FROM yearmonth
        GROUP BY customerid
    )
    UNION ALL
    SELECT
        ym.customerid,
        ym.date::INT,
        c.running_credit + GREATEST(ym.consumption - 700, 0) AS running_credit
    FROM yearmonth ym
    JOIN loyalty c
        ON ym.customerid = c.customerid
        AND ym.date::INT = CASE
            WHEN c.date % 100 = 12 THEN (c.date / 100 + 1) * 100 + 1
            ELSE c.date + 1
        END
)
SELECT customerid, MIN(date) AS first_milestone_month
FROM loyalty
WHERE running_credit > 1000
GROUP BY customerid;
\end{lstlisting}

\subsection{Warm-up Task}
\label{app:warmup}

Participants completed the following warm-up task before beginning the
four study tasks. The warm-up familiarized participants with the task
format and study interface; data from this task were excluded from the
analysis.

\paragraph{Task Requirements.}
What product names does each gas station offer?

\paragraph{Buggy SQL Query.}

\begin{lstlisting}
SELECT s.gasstationid, s.productid 
FROM station_catalog s LEFT JOIN products p ON s.productid = p.productid;
\end{lstlisting}

\section{Pre-Task Survey}
\label{app:pre-survey}

\section{Post-Task Survey}
\label{app:post-survey}

We reproduce below the post-task survey items used in our analysis.
Items not analyzed in this paper, as well as administrative fields, are omitted.


\section{Survey and Interview Materials}
\label{app:survey-interview-materials}

\subsection{NASA-TLX}
\label{app:survey-llm-tasks}

The following questions were adapted from the NASA-TLX~\cite{HART1988139} and assesses workload across the two study conditions and participants' perceptions of LLM responses during the study.

\subsubsection{LLM+Search Condition}
\begin{enumerate}

    \item How mentally demanding were the SQL debugging tasks with
    access to a large language model (LLM)?

    \textit{Response:} 7-point scale,
    1 = Not demanding; 7 = Very demanding.

    \item How rushed or pressured did you feel while working on the
    tasks with access to an LLM?

    \textit{Response:} 7-point scale,
    1 = Not rushed; 7 = Very rushed.

    \item How much effort did you need to put into debugging the SQL
    queries with access to an LLM?

    \textit{Response:} 7-point scale,
    1 = No effort; 7 = High effort.

    \item How frustrated, stressed, or annoyed did you feel while
    debugging the queries with access to an LLM?

    \textit{Response:} 7-point scale,
    1 = Not frustrated; 7 = Very frustrated.

    \item How satisfied were you with how you handled the debugging
    tasks with access to an LLM?

    \textit{Response:} 7-point scale,
    1 = Not satisfied; 7 = Very satisfied.

\end{enumerate}

\subsubsection{Search-only Condition}
\label{app:survey-no-llm-tasks}

\begin{enumerate}

    \item How mentally demanding were the SQL debugging tasks without
    access to an LLM?

    \textit{Response:} 7-point scale,
    1 = Not demanding; 7 = Very demanding.

    \item How rushed or pressured did you feel while working on the
    tasks without access to an LLM?

    \textit{Response:} 7-point scale,
    1 = Not rushed; 7 = Very rushed.

    \item How much effort did you need to put into debugging the SQL
    queries without access to an LLM?

    \textit{Response:} 7-point scale,
    1 = No effort; 7 = High effort.

    \item How frustrated, stressed, or annoyed did you feel while
    debugging the queries without access to an LLM?

    \textit{Response:} 7-point scale,
    1 = Not frustrated; 7 = Very frustrated.

    \item How satisfied were you with how you handled the debugging
    task without access to an LLM?

    \textit{Response:} 7-point scale,
    1 = Not satisfied; 7 = Very satisfied.

\end{enumerate}

\subsection{In-Study LLM Response Satisfaction}
\label{app:survey-llm-study}

\begin{enumerate}

    \item How satisfied were you with the LLM's responses during this
    study?

    \textit{Response:} 7-point scale,
    1 = Not satisfied; 7 = Very satisfied.

    \item How much did you trust the LLM's responses during this
    study?

    \textit{Response:} 7-point scale,
    1 = No trust; 7 = High trust.

    \item How often did you fully understand the LLM's responses
    during this study?

    \textit{Response:} 7-point scale,
    1 = Not often; 7 = Very often.

    \item After using the LLM during this study, how confident are you
    that you could apply the same fix to a similar SQL problem on
    your own?

    \textit{Response:} 7-point scale,
    1 = Not confident; 7 = Very confident.

\end{enumerate}

\subsection{Self-Report LLM-Use, Overuse, and Dependence}
\label{app:survey-general-llm}

We adapted four self-report questions related to LLM use, overuse, and dependence from prior work by Li et al., which reported positive correlations between these items and scores on the 18-item LLM Dependence Scale~\cite{DBLP:journals/ijinfoman/LiZWW25}. We used the four-item measure to reduce participant burden.

\begin{enumerate}

    \item How frequently have you used LLMs in the past week?

    \textit{Response options:}
    \begin{itemize}
        \item Never used
        \item Rarely used (no more than once a week)
        \item Occasionally used (about 2--3 times a week)
        \item Frequently used (about 4--5 times a week)
        \item Very frequently used (more than 6 times a week)
    \end{itemize}

    \item How much time do you typically spend using LLMs each
    session?

    \textit{Response options:}
    \begin{itemize}
        \item Less than 30 minutes
        \item 30 minutes to 1 hour
        \item 1--2 hours
        \item 2--5 hours
        \item More than 5 hours
    \end{itemize}

    \item Do you believe you overuse LLMs?

    \textit{Response:} 7-point scale,
    1 = Strongly disagree; 7 = Strongly agree.

    \item Do you think you are dependent on LLMs?

    \textit{Response:} 7-point scale,
    1 = Strongly disagree; 7 = Strongly agree.

\end{enumerate}

\subsection{SQL \& LLM Experience}
\label{app:survey-sql-experience}

\begin{enumerate}

    \item How comfortable are you debugging SQL queries?

    \textit{Response:} 7-point scale,
    1 = Not comfortable; 7 = Very comfortable.

    \item How often do you write or debug SQL queries?

    \textit{Response options:}
    Daily; Weekly; Monthly; Quarterly; Annually; Other.

    \item How many years have you actively used SQL
    (coursework, work, or projects)?

    \textit{Response options:}
    \begin{itemize}
        \item Less than 1 year
        \item 1--2 years
        \item 3--5 years
        \item 6--10 years
        \item More than 10 years
    \end{itemize}

    \item In what contexts have you used SQL?
    \textit{(Select all that apply.)}

    \textit{Response options:}
    \begin{itemize}
        \item Coursework or class projects
        \item Research projects
        \item Work or industry projects
        \item Personal projects
        \item Other
    \end{itemize}

    \item How frequently do you use large language model (LLM) tools
    (e.g., ChatGPT, Claude, Gemini) in your daily work or studies?
    \textit{Response options:}
    Daily; Weekly; Monthly; Quarterly; Annually; Other.

\end{enumerate}

\subsection{Demographic Information}
\label{app:survey-demographics}

\begin{enumerate}

    \item Age range

    \textit{Response options:}
    18--24; 25--34; 35--44; 45--54; 55--64; 65+;
    Prefer not to say.

    \item Gender identity

    \textit{Response options:}
    Female; Male; Non-binary / gender non-conforming;
    Prefer not to say; Other.

\end{enumerate}

\subsection{Post-Task Interview Protocol}
\label{app:interview-protocol}

The following questions served as a pool of potential follow-up questions for the post-task interview. Questions were asked as appropriate depending on the participant's responses and available time.

\textit{Would you be open to answering a few follow-up questions?}
[If yes, proceed. If no, skip to closing.]
\begin{itemize}
    \item \textbf{[Helpful Tools]} What did you find most helpful in debugging?
    \item \textbf{[LLM Positives]} What kinds of things was the LLM good at during these tasks?
    \item \textbf{[LLM Negatives]} While using the LLM during this study, were there things you expected it to help with that it did not?
    \item \textbf{[Experiment Challenges]} What did you find most challenging while debugging the SQL queries?
    \item \textbf{[LLM Use Decision]} Can you walk me through your decision about whether or not to use ChatGPT during the tasks?
    \item \textbf{[Instructor Guidelines]} \textit{If the participant is a student:} Have you received guidelines from your class instructors on how to use LLMs?
    \item \textbf{[Verification Strategies]} If you had more time, is there something else you would have done to help you verify query correctness?
    \item \textbf{[LLM Trustworthiness]} Do you have any other thoughts regarding how trustworthy and useful LLMs are for SQL debugging?
    \item \textbf{[LLM Performance]} What did you think about how the LLM performed during these tasks?
    \item \textbf{[Open-Ended]} Are there any other thoughts you would like to share?
\end{itemize}

\section{Coding Materials}
\label{app:coding}

The coding schemes below summarize the codes used during analysis. Prompt-level codes were briefly summarized in the analyses reported in Section~\ref{sec:results_llm}. Transcript and video codes served as lower-level analytic scaffolding; the higher-level categories and patterns reported in the findings were developed through subsequent interpretation across coded data.

\subsection{Prompt Type Coding}
\label{app:prompt-type}

We coded each LLM prompt according to the type of assistance requested (see Table~\ref{tab:prompt-type-codebook}). Prompts containing a request were assigned one request type. Prompts that did not request an action were coded as statements.

\begin{table*}[t]
\centering
\small
\caption{Prompt type coding scheme.}
\label{tab:prompt-type-codebook}
\Description{The table contains a coding scheme that defines nine request types and two statement types. Request codes distinguish prompts that ask for explanation, error detection, correction, simplification, verification, artifact generation, strategy, SQL syntax identification, or review. Statement codes distinguish reports of observed behavior from corrections to prior information or task requirements.}

\begin{tabularx}{\textwidth}{@{}p{0.12\textwidth}p{0.15\textwidth}X@{}}
\toprule
\textbf{Category} & \textbf{Code} & \textbf{Definition} \\
\midrule

Request
& Explain
& Asks the LLM to explain a SQL query or fragment, SQL syntax,
the task requirements, or a combination of these. Includes implicit
requests such as ``What does this do?'' \\
\cmidrule(lr){2-3}

& Detect
& Asks the LLM to identify an error in the original SQL query. \\
\cmidrule(lr){2-3}

& Fix
& Asks the LLM to correct or write SQL to produce a desired behavior,
including follow-up requests that clarify the intended behavior. \\
\cmidrule(lr){2-3}

& Simplify
& Asks the LLM for a simpler or alternative solution to the SQL task. \\
\cmidrule(lr){2-3}

& Verify
& Asks the LLM to assess whether a candidate solution is correct, or
presents a candidate solution for assessment. \\
\cmidrule(lr){2-3}

& Generate
& Asks the LLM to create a supporting artifact for reasoning about the
task without directly correcting the SQL, such as examples or a formal
representation. \\
\cmidrule(lr){2-3}

& Strategize
& Asks the LLM to suggest an approach or possible next steps. \\
\cmidrule(lr){2-3}

& Identify (Syntax)
& Asks how to perform a specific operation in SQL or whether an
appropriate SQL construct exists. \\
\cmidrule(lr){2-3}

& Review
& Presents an interpretation or hypothesis for the LLM to assess,
including cases without an explicit request. \\

\midrule

Statement
& Report
& Reports an observed behavior or result without explicitly asking
the LLM to take an action. \\
\cmidrule(lr){2-3}

& Correction
& Provides a correction to the task requirements or to information
previously given to the LLM. \\

\bottomrule
\end{tabularx}
\end{table*}

\subsection{Automation Coding}
\label{app:automation-reasoning}

In addition to prompt type, we coded each prompt along the amount of debugging work delegated to the LLM (see Table~\ref{tab:automation-reasoning-codebook}).


\begin{table*}[t]
\centering
\small
\caption{Automation coding schemes.}
\label{tab:automation-reasoning-codebook}
\Description{Table of a coding scheme for automation that has three levels of delegated debugging work. Low supports understanding without substantive debugging, moderate delegates a bounded step within a participant-identified approach, and high delegates substantive diagnosis, repair, or other debugging work to the large language model.}

\begin{tabularx}{\textwidth}{@{}p{0.20\textwidth}p{0.18\textwidth}X@{}}
\toprule
\textbf{Dimension} & \textbf{Code} & \textbf{Definition} \\
\midrule

Automation
& Low
& The LLM supports the participant's understanding of the SQL or task
without performing a substantive part of the debugging process. \\
\cmidrule(lr){2-3}

& Moderate
& The LLM performs a bounded step that advances a
participant-identified approach without taking over the overall
diagnosis or repair. \\
\cmidrule(lr){2-3}

& High
& The LLM is asked to perform diagnosis, repair, or another substantive
part of the debugging process. \\

\bottomrule
\end{tabularx}
\end{table*}

\subsection{Transcript Coding Summary}
\label{app:transcript-coding}

Transcript coding was used to identify recurring debugging and
reasoning practices and participants' perceptions of the LLM. Table~\ref{tab:transcript-codes} summarizes the lower-level codes used during this analysis. These codes supported subsequent interpretive analysis; they were not treated as the final themes reported in the paper.

\begin{table*}[t]
\centering
\small
\caption{Summary of lower-level transcript codes.}
\label{tab:transcript-codes}
\Description{Table of lower-level transcript codes that are organized into four families: debugging, reasoning, verification, and large language model perceptions. The families capture how participants inspected and diagnosed problems, reasoned about task requirements and SQL behavior, evaluated candidate solutions, and described the utility, reliability, and role of large language model assistance.}

\begin{tabularx}{\textwidth}{@{}p{0.21\textwidth}p{0.28\textwidth}X@{}}
\toprule
\textbf{Code Family} & \textbf{Codes} & \textbf{Description} \\
\midrule

Debugging strategies
& Inspect schema; Inspect query output; Inspect intermediate output;
Annotate SQL with comments
& Actions participants described or performed to inspect the query,
data, or intermediate behavior while diagnosing a problem. \\
\midrule

Reasoning strategies
& NL--SQL mapping; Example-based reasoning
& Ways participants reasoned about the relationship between task
requirements, SQL behavior, and concrete examples. \\
\midrule

Verification strategies
& Test cases; Spot check; Result comparison; Inspect query output;
Ask LLM; Comprehension walk-through; LLM corroboration
& Approaches participants used or described for evaluating whether a
candidate solution was correct. \\
\midrule

LLM perceptions
& Utility; Trust; Distrust; Reliance; Conditional reliability;
Independence
& Statements expressing participants' perceptions of the usefulness
and reliability of LLM assistance and the role they wanted it to play
in their own work. \\

\bottomrule
\end{tabularx}
\end{table*}

\subsection{Video Coding Summary}
\label{app:video-coding}

Video coding captured participants' observable activity during each task (see Table~\ref{tab:video-codes}). The original coding scheme distinguished finer-grained variants of several activities (e.g., inspecting the output of the original query versus a candidate query). For the interaction-pattern analysis, these lower-level activities were subsequently consolidated into the higher-level states of Query Orienting, Query Fixing, and Query Verification described in Section~\ref{sec:results_activties}.

\begin{table*}[t]
\centering
\small
\caption{Summary of activities identified during video coding.}
\label{tab:video-codes}
\Description{Table of a coding scheme for the video codebook that includes eleven observable activities spanning inspection, reasoning, editing, and tool use. Activities include reviewing task requirements, Structured Query Language, query output, schemas, or data; using scratch paper; formulating hypotheses; editing queries; prompting or interpreting large language model output; and using web searches or external resources.}

\begin{tabularx}{\textwidth}{@{}p{0.24\textwidth}X@{}}
\toprule
\textbf{Activity} & \textbf{Description} \\
\midrule

Task-requirement walkthrough
& Reading or revisiting the natural-language task requirements,
including silently tracing or highlighting them. \\
\midrule

SQL walkthrough
& Reading or tracing the original SQL or a candidate solution.
Candidate-query walkthroughs were distinguished during coding by
whether participants appeared to be diagnosing or verifying the query. \\
\midrule

Inspecting query output
& Examining results from the original query, a candidate solution, or
an exploratory/intermediate query. \\
\midrule

Inspecting schema or data
& Consulting table schemas, column information, or raw database data. \\
\midrule

Using scratch paper
& Writing on paper to reason about the task, query, schema, or data. \\
\midrule

Hypothesis formulation
& Articulating a hypothesis about the source of the error or a
possible repair. \\
\midrule

Editing SQL
& Making substantive edits toward a candidate solution or modifying a
query for exploratory purposes. \\
\midrule

Prompting the LLM
& Composing or submitting a prompt to the LLM. \\
\midrule

Reading/interpreting LLM output
& Reading, processing, or verbalizing an interpretation of an LLM
response. \\
\midrule

Web search
& Entering or reviewing a search-engine query. \\
\midrule

Reading external resources
& Reading documentation or another external reference reached through
search. \\

\bottomrule
\end{tabularx}
\end{table*}

\subsection{LLM-Assisted Coding Review}
\label{app:llm-coding-review}

Two general-purpose LLMs independently reviewed the final prompt-level coding using the code definitions in Tables~\ref{tab:prompt-type-codebook} and \ref{tab:automation-reasoning-codebook}. For each prompt, the models
were provided with the participant's prompt text, the assigned code,
coder notes when available, and the original task materials when needed
for context.

For each coding dimension, the model returned
\textit{Keep}, \textit{Change}, or \textit{Unclear}.
Models were instructed to use \textit{Keep} when the existing label was
consistent with the codebook, rather than changing a label merely
because another interpretation was plausible, and to use
\textit{Unclear} when the available evidence was insufficient.
Each judgment included a brief rationale grounded in the prompt text.

\paragraph{Automation.}
Models were instructed to code based on the debugging responsibility
delegated to the LLM rather than the topic of the prompt. For prompts
containing multiple requests, the highest applicable automation level
was used. Evaluating a participant-developed candidate solution or
supporting a participant-identified approach was treated as Moderate,
whereas broadly diagnosing the original query or generating or repairing
the target solution was treated as High.


\paragraph{Prompt Type.}
Models used the prompt-type definitions in
Table~\ref{tab:prompt-type-codebook}. When a prompt contained multiple
requests or statements, they were instructed to select the type that
best represented the prompt's primary communicative intent. Task
materials were consulted when needed to distinguish the original query
from a participant-generated candidate query.

\section{LLM Evaluation}
\label{app:llm-eval}

GPT-5.5 achieved 100\% accuracy on Tasks A--C in both debugging conditions, but 0\% on Task D (see Table~\ref{tab:llm_performance}). A brief review suggests that GPT-5.5 may have interpreted the ``consecutive months'' requirement more broadly than the gold SQL by explicitly accounting for gaps in monthly records and restarting accumulation afterward. This difference in interpretation appears to have contributed to a higher output row count than the gold SQL. GPT-5.4 was similarly strong, and more consistent with our gold SQL, achieving 90--100\% accuracy without the schema across all four tasks. GPT-5.4-mini achieved high accuracy on the Universal Quantification tasks (A and B) but was less reliable on the Recursive tasks (C and D). 

    \begin{table}[t]
    \centering
    \caption{Execution accuracy (\%) across 10 runs for each model--prompt--task combination. \textit{Debug} provides the task requirements and candidate query; \textit{Generate} provides only the task requirements. \textit{+Schema} additionally provides the database schema. We included generation without the schema because, during pilot testing with benchmark queries, models sometimes produced correct queries without being given schema information; no model did so on the study tasks. A response was counted as correct when its query returned the same rows as the reference query, following execution-based evaluation used in Spider 2.0 \cite{DBLP:conf/iclr/LeiCYCSSSGHYZX025}.}
    \label{tab:llm_performance}
    \Description{Table of execution accuracy for three models across four tasks and four prompting conditions. Debugging accuracy was generally high, while generation without the schema was zero percent throughout. Generation with the schema was high for Tasks A and B but low for Tasks C and D across both models.}

    \begin{tabular}{llrrrr}
    \toprule
    Model & Task & Debug & Debug+Schema & Generate & Generate+Schema \\
    \midrule
    GPT-5.4
        & A & 100 & 70  & 0 & 100 \\
        & B & 100 & 90  & 0 & 100 \\
        & C & 100 & 100 & 0 & 10  \\
        & D & 90  & 60  & 0 & 0   \\
    \addlinespace
    
    GPT-5.4-mini
        & A & 90  & 100 & 0 & 90 \\
        & B & 100 & 70  & 0 & 90 \\
        & C & 50  & 40  & 0 & 0  \\
        & D & 10  & 10  & 0 & 0  \\
    \addlinespace
    
    GPT-5.5
        & A & 100 & 100 & 0 & 100 \\
        & B & 100 & 100 & 0 & 100 \\
        & C & 100 & 100 & 0 & 10  \\
        & D & 0   & 0   & 0 & 0   \\
    \addlinespace
    
    \bottomrule
    \end{tabular}
    \end{table}

Query generation showed a different pattern. With the schema, generation achieved 90--100\% execution accuracy on the Universal Quantification tasks but only 0--10\% on the Recursive tasks. This contrast suggests that the models matched our reference-query behavior more reliably when repairing recursive queries than when generating them from scratch.
\end{document}